\documentclass{aa}  
\usepackage{multicol}
\usepackage{graphicx}
\usepackage{txfonts}
\usepackage{pdfpages}
\usepackage{bm}
\usepackage{balance}
\begin{document}

   \title{Fragmentation of star-forming filaments in the X-shape Nebula of the California  molecular cloud}


   \author{Guo-Yin Zhang
          \inst{1,2,3}
           \and
          Ph.~Andr\'e\inst{2}
          \and
          A.~Men'shchikov\inst{2}
          \and
          Ke Wang\inst{4,5}
          }
   \institute{National Astronomical Observatories, Chinese Academy of Sciences, Beijing 100101, PR China\\
              \email{zgyin@nao.cas.cn}
         \and Laboratoire d'Astrophysique (AIM), CEA/DRF, CNRS, Universit\'e Paris-Saclay, Universit\'e Paris Diderot, Sorbonne Paris Cit\'e, 91191 Gif-sur-Yvette, France\\
             \email{philippe.andre@cea.fr; alexander.menshchikov@cea.fr}
             \and 
             University of Chinese Academy of Sciences, Beijing 100049, PR China
             \and
             Kavli Institute for Astronomy and Astrophysics, Peking University, 5 Yiheyuan Road, Haidian District, Beijing 100871, PR China\\
             \email{kwang.astro@pku.edu.cn}
             \and
             European Southern Observatory (ESO) Headquarters, Karl-Schwarzschild-Str. 2, 85748 Garching bei M\"{u}nchen, Germany
             }

   \date{Received 13 February 2020; accepted  27 July 2020}

 
  \abstract
   {Dense molecular filaments are central to the star formation process, but
   the detailed manner in which they fragment into prestellar cores is not yet well understood.}
 {Here, we investigate the fragmentation properties and dynamical state of several star-forming filaments
 in the X-shape Nebula region of the California  molecular cloud, in an effort to shed some light
 on this issue. }
   {We used multi-wavelength far-infrared images from {\it Herschel} and 
  the \textsl{getsources} and \textsl{getfilaments} extraction methods to 
  identify dense cores and filaments in the region and derive their basic properties.
 We also used  a map of $\rm ^{13}CO (2-1)$ emission from the Arizona 10m Submillimeter Telescope (SMT) 
 to  constrain the dynamical state of the filaments.  
   }
   {We identified 10 filaments with aspect ratios $AR >4$ and column density contrasts $C>0.5$, 
   as well as 57 dense cores, including 2 protostellar cores, 20 robust prestellar cores, 11 candidate prestellar cores, 
   and 24 unbound starless cores. 
   All 10 filaments have roughly the same deconvolved FWHM width, 
   with a median value $0.12\pm 0.03$ pc, 
   independently of their column densities ranging from $< 10^{21}\, {\rm cm}^{-2}$ to  $>10^{22}\, {\rm cm}^{-2}$.
   Two star-forming filaments (\# 8 and \# 10) stand out in that they 
   harbor quasi-periodic chains of dense cores with a typical projected core 
   spacing of $\sim$0.15 pc. 
   These two filaments have thermally supercritical line masses and are not static. 
   Filament~8 exhibits a prominent transverse velocity gradient, suggesting that it 
   is accreting gas from the parent cloud gas reservoir at an estimated rate of 
   $\sim 40\pm10\ M_\sun\ \rm Myr^{-1}\ pc^{-1}$. 
   Filament~10 includes two embedded protostars with outflows and 
   is likely at a somewhat later evolutionary stage than filament~8.
   In both cases, the observed (projected) core spacing is similar to the filament width and significantly shorter than the canonical separation of 
   $\sim \,$4 times the filament width predicted by classical cylinder fragmentation theory. 
   Projection effects are unlikely to explain this discrepancy.  
   We suggest that continuous accretion of gas onto the two star-forming filaments, as well as geometrical bending of the filaments, 
   may account for the observed core spacing.
}
{ Our findings suggest that the characteristic fragmentation lengthscale of molecular filaments is 
quite sensitive to external perturbations from the parent cloud, such as gravitational accretion of ambient material.} 
   \maketitle
%
\section{Introduction}
 The early phases of star formation are still poorly understood. 
  One of the most pressing open questions is 
  how filaments fragment into cores 
  at the earliest evolutionary stages. 
  Addressing this issue is crucial to our understanding of 
  the initial conditions of star formation (cf. \citealt{Andre2010}).
  
  Observations with the {\it Herschel} space observatory have revealed that filaments are 
  truly ubiquitous in the cold interstellar medium (ISM), with lengths ranging from $\rm \sim pc$ in nearby 
  molecular clouds to $\rm \sim 10^{2}\ pc$ in the Galactic plane \citep{Men2010,Wang2015}.
  Most ( > 75 \%) prestellar cores lie inside filaments with column densities 
  $N_{\rm H_{2}}^{\rm fil} \gtrsim 7\times 10^{21}\rm\ cm^{-2}$ \citep{Konyves2015}, 
  implying that filaments play a key role in the star formation process \citep{Andre2014}. 
  The typical inner width of filaments in nearby molecular clouds is $\rm \sim 0.1\ pc$ 
  \citep{Arzoumanian2011, Arzoumanian2019}, but the origin of this characteristic width is 
  still a controversial topic.
  The typical value may come from supersonic turbulence in the ISM \citep{Pudritz2013, Federrath2016}, 
  or the balance of quasi-equilibrium structure
  with ambient ISM pressure \citep{Fischera2012}.
  Filaments serve as highly efficient routes for feeding material into 
  star-forming cores \citep{Andre2014}. The material around filaments is not 
  at all static. For instance, a pronounced transverse velocity gradient provides 
  good kinematic evidence for accretion flows of ambient gas material into the B211/B213 filament 
  of the Taurus molecular cloud (MC) 
  \citep{Palmeirim2013,Shimajiri2019}. 
  Other dense filaments such as infrared dark clouds or the Serpens South filament
  exhibit both transverse and longitudinal velocity gradients, suggesting that, gas is not 
  only accreted by, but also flowing along these filaments \citep[e.g.][]{Kirk2013}. 
  In most cases, the transverse gradients appear to dominate over the longitudinal gradients \citep[][]{Dhabal2018}. 
  In the Serpens South case, for instance, the mass flow rate along the filament is estimated to be 
  only $\sim $1/4 of the accretion rate in the transverse direction \citep{Kirk2013}.
  The critical line-mass of a cylindrical isothermal filament in hydrostatic equilibrium is 
  $M_{\rm line, crit} \equiv 2\, c_{\rm s}^2/G$, where $c_{\rm s}$ is the sound speed and $G$ the gravitational constant, 
  or $\sim $16 $M_{ \odot }\ \rm pc^{-1}$ for a gas temperature of 10 K \citep[e.g.][]{Ostriker1964,Inutsuka1997}.  
  Recently, \citet{Arzoumanian2019} divided observed filaments into three families according their 
  line-mass $M_{\rm line}$: thermally supercritical filaments ($M_{\rm line}  \gtrsim 2\, M_{\rm line, crit}$),
  transcritical filaments 
  ($0.5\, M_{\rm line, crit} \lesssim M_{\rm line}  \lesssim 2\, M_{\rm line, crit}$),
  thermally subcritical filaments 
  ($M_{\rm line, crit} \lesssim 0.5\, M_{\rm line, crit}$). 
  For an infinitely long cylindrical filament in hydrostatic equilibrium, 
  core spacing is predicted to be $\sim$ 4 $\times$ the filament diameter 
  by linear fragmentation models \citep[e.g.][]{Inutsuka1992},
  or $\sim \,$0.4 pc taking the typical filament width into account.
  However, this does not match the actual core spacing found in observations 
  which, at least on small scales, appears to be dominated by typical Jeans-like fragmentation
  \citep[e.g.][]{Kainulainen2013,Kainulainen2017,Konyves2020,Ladjelate2020}.
 Here, based on our findings in the California MC, 
 we propose that this discrepancy results from the fact that star-forming filaments 
 are not isolated but accrete fresh matter from their parent clouds at the same time 
 as they fragment into cores.
  
\begin{figure}
 \centering
\includegraphics[angle=90, width=0.5 \textwidth]{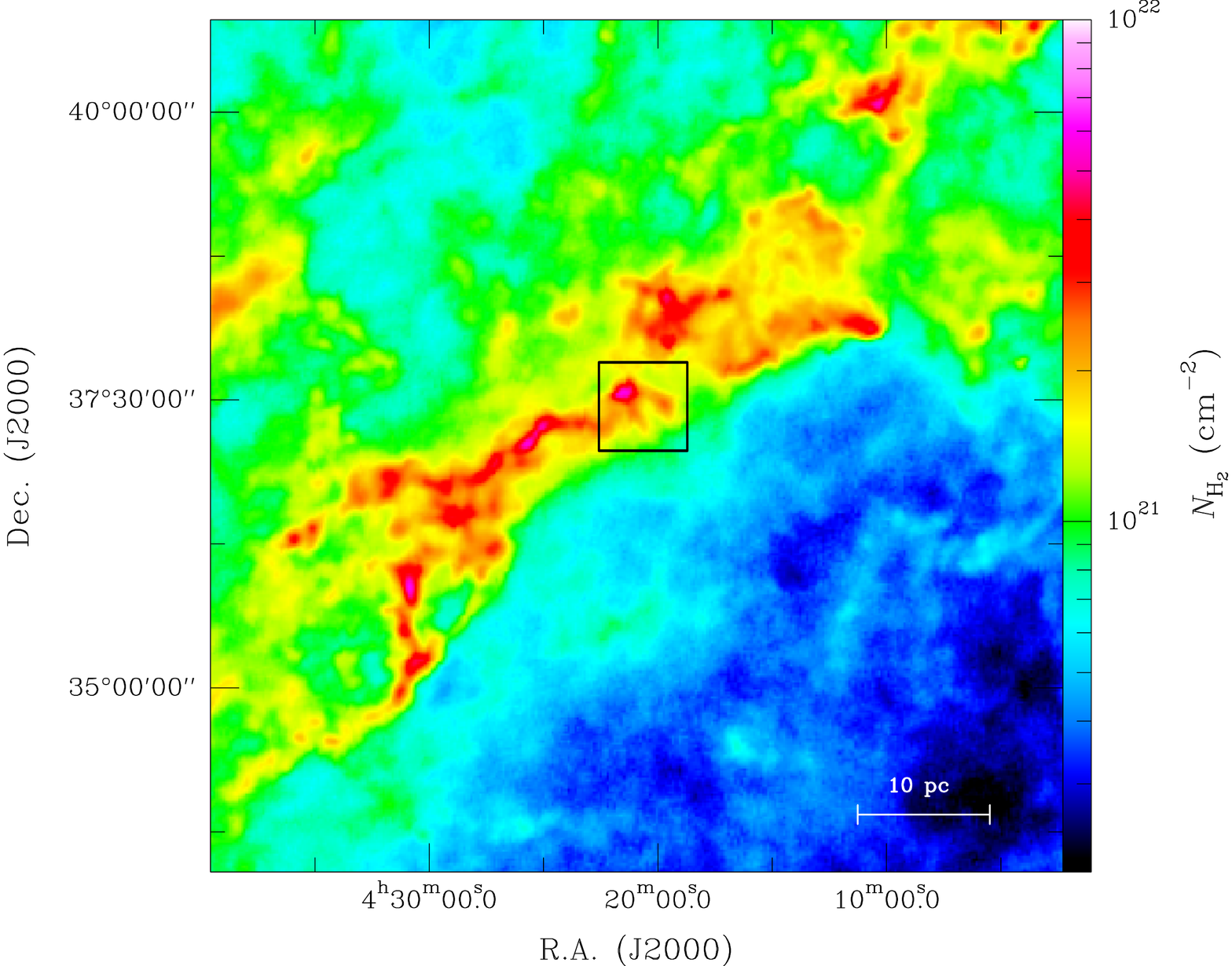}
\caption{
Large-scale column density map of the California MC derived from 
 {\it Planck} $850\, \mu$m optical-depth data ($5\arcmin $ resolution).
The location of the X-shape Nebula region is marked by black square. 
} 
\label{californiaC}
\end{figure}

  Star formation activity is generally found only in high-extinction parts of MCs,
  but the mass of high-extinction material with $A_{K}>1.0$ mag in the California MC 
  is only 10\% of that in OrionA~MC \citep{Lada2009}, and the star formation rate 
  is accordingly much lower.
  The number of young stellar objects (YSOs), which may be taken as an indicator of star-formation activity, 
  is only 177/2980$\sim$ 6\% of that observed in the Orion~A MC \citep{Lada2017,Groschedl2019}.
 Only one B-type main-sequence star 
 is found in the south-eastern part of the California MC and is associated 
 with relatively intense star formation in a local region around it \citep{Andrews2008}. 
 The global star formation efficiency estimated by 
   \citet{Zhang2018} for the California MC is only $\sim 1 \%$, 
   which is half of the typical value ($\sim 2 \%$) 
  in the molecular clouds of the Milky Way \citep{Evans1991}.  
  The California MC is therefore an ideal place to study star formation at early stages. 
  The X-shape region lies at the center of the California MC (see Fig.~\ref{californiaC}).
  The distance to the California MC as estimated from {\it Gaia} DR2 \citep{Gaia2018} stellar parallaxes  
  is $500 \pm 7$ pc according to \citet{Yan2019} and $470 \pm 2 \pm 24$ pc according to \citet{Zucker2019}, 
  which is slightly farther than both the distance ($450\pm23$ pc) estimated by \citet{Lada2009}
  through comparison of foreground star counts with Galactic models  and 
  the distance ($410\pm 41$ pc) estimated by \citet{Schlafly2014} with stars from the PanSTARRS-1 survey. 
  Different distance measurement methods bring uncertainties of about 10 \% in size and  20 \% in mass 
  for the California MC.
  In this study, we adopt a distance of 500 pc.
  The most distinctive feature of this region is that it resembles an 'X'.
  Two low-density filaments meet at the north dense hub region 
  and extend to the southeast and southwest 
  with an intersection angle of $\sim 60^{\circ }$ in the plane of sky. 
  Small longitudinal velocity gradients of 0.1 and 0.2 $\rm km\ s^{-1}\ pc^{-1}$ 
  are measured along the southeast and southwest filaments \citep{Imara2017}.  
  The hub region harbors at least two YSOs. 
  One is a Class II object and the other is a Class I  source \citep{Harvey2013, Broekhoven2014}. 
  These two YSOs may be the driving sources of a low-mass low-velocity outflow \citep{Imara2017}. 
  
 The outline of the present paper is as follows. 
 In Sect. \ref{sec:observations}, 
  we describe the {\it Herschel} submillimeter dust emission data 
  and SMT 10m molecular line observations of the X-shape region.  
  Data analysis and results 
  are presented in Sect. \ref{sec:analysis}. In Sect.~\ref{sec:discussion}, 
  we discuss  evidence of accretion onto an early-stage filament, 
  as well as the detailed fragmentation properties of the same filament.
  We summarize our conclusions in Sect.~\ref{sec:conclusions}.
 \begin{figure*}
\begin{minipage}{0.49 \linewidth}
   \centering
\vspace{-0.7cm}
\includegraphics[width=0.8 \textwidth]{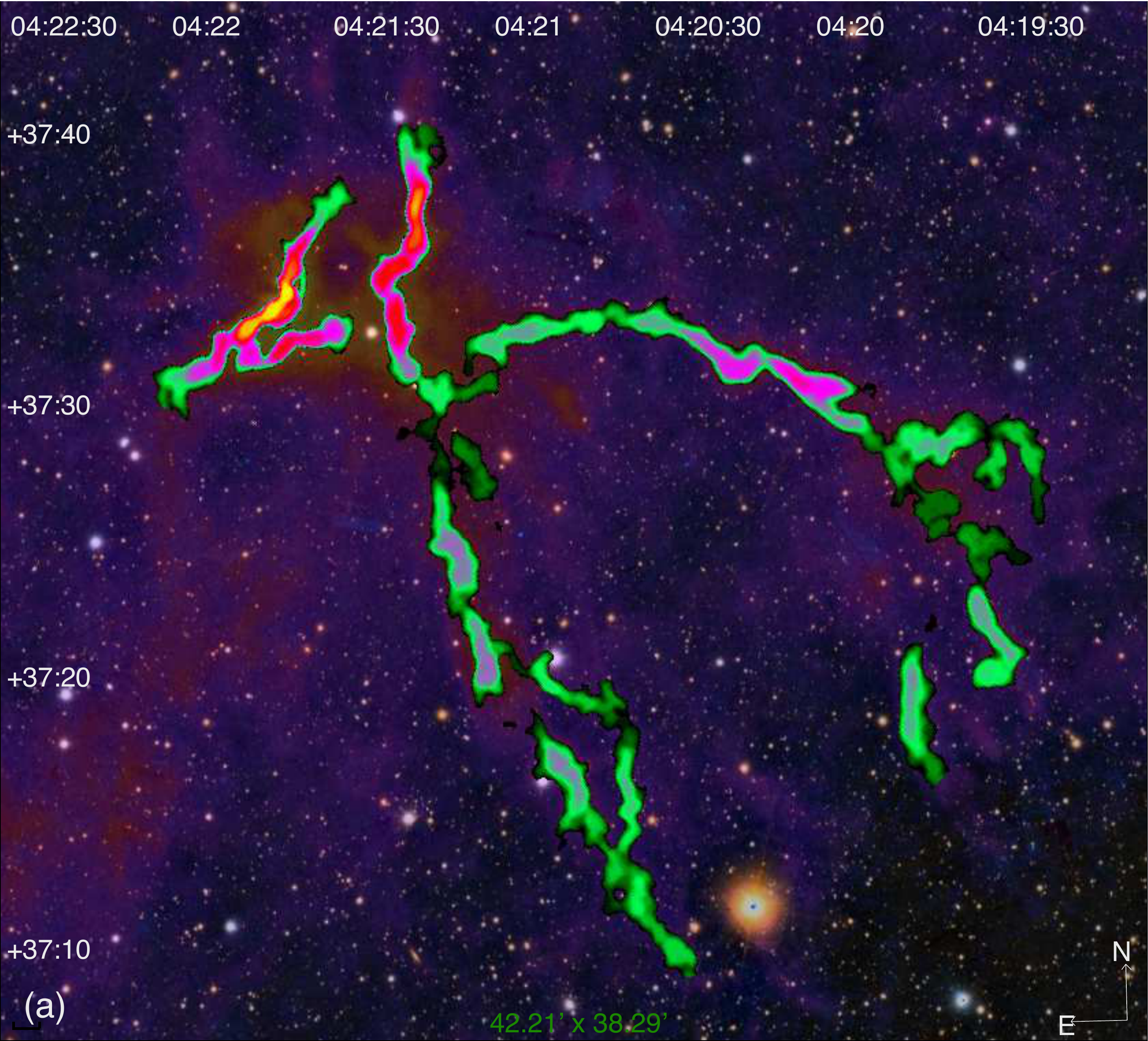}
\end{minipage}
\begin{minipage}{0.49 \linewidth}
\centering
\includegraphics[width=1.0 \textwidth]{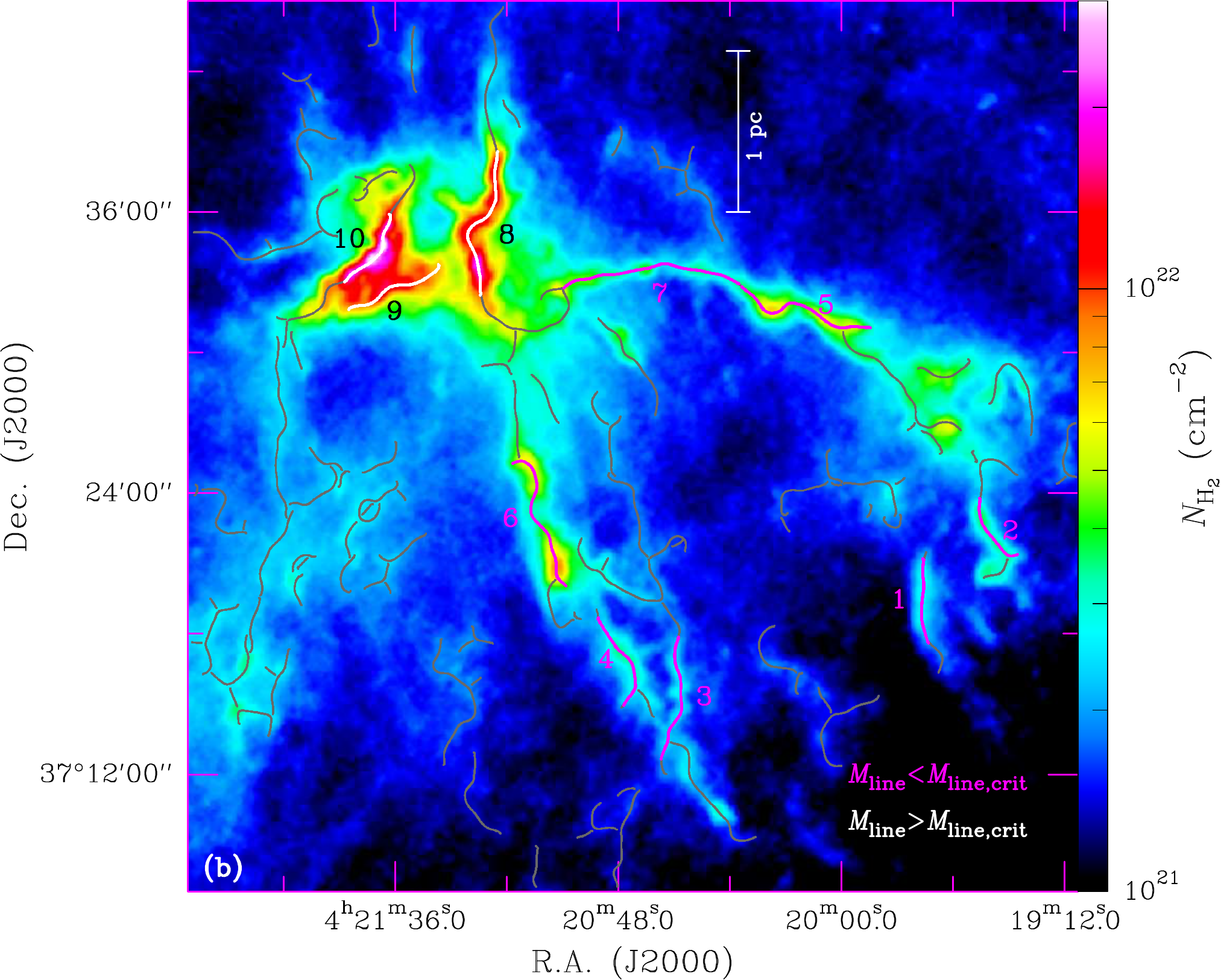}
\end{minipage}
\caption{Blow-up column density views of the X-shape Nebula region in the California MC.
Panel (A): Filtered version of the high-resolution (18.2\arcsec) column density map 
showing partially-reconstructed filaments with 
transverse angular scales only up to $\sim 100${\arcsec} 
from \textsl{getfilaments} \citep[cf. Sect.\,2.4.3 in][]{Men2013}.
The background image shows field stars from the Pan-STARRS optical survey  
 \citep{Chambers2016}. 
Panel (B): Unfiltered high-resolution column density map with the network of 
filamentary structures identified by \textsl{getfilaments} overlaid in  grey. 
The crests of the ten filaments selected in Sect.~\ref{sec:fil_extraction} are 
highlighted in white (for thermally supercritical filaments with $M_{\rm line}>M_{\rm line,crit}$)
or magenta (for subcritical filaments with $M_{\rm line}<M_{\rm line,crit}$).
}
         \label{map}
\end{figure*} 
\section{Observations and data reduction}\label{sec:observations}
\subsection{Herschel dust continuum data}
The {\it Herschel} imaging observations of the California MC \citep{Harvey2013},
include PACS 70 and 160 $\mu$m \citep{Poglitsch2010} and SPIRE 250, 350 and 500 $\mu$m \citep{Griffin2010} data. 
The beam sizes of the PACS data at 70 and 160 $\mu$m are 8.4 and $13.5${\arcsec}, respectively. 
The beam sizes of the SPIRE data at 250, 350 and 500 $\mu$m are 18.2, 24.9 and $36.3${\arcsec}, respectively. 
The SPIRE/PACS parallel-mode was used to make scan maps with  speed of $60${\arcsec} s$^{-1}$. 
We downloaded the {\it Herschel} maps from the NASA/IPAC Infrared Science 
Archive\footnote{https://irsa.ipac.caltech.edu/data/Herschel/ACMC/} 
and extracted a sub-field of  about 40{\arcmin}$\times $40{\arcmin} around the X-shape Nebula. 
A close-up map of the X-shape Nebula is shown in Fig.~\ref{map}. 
The location of the X-shape Nebula is marked in the large-scale map of the California MC shown in Fig.~\ref{californiaC}.
Zero-level offsets were derived for the {\it Herschel} maps 
from comparison with {\it Planck} and IRAS data (cf. Bracco et al., in prep.; \citealp{Bernard2010}).
They are 19.7, 26.5, 15.0,  and 5.5   $\rm MJy\ sr^{-1}$
at 160, 250, 350, and 500 $\mu$m, respectively. 
Pixel-by-pixel SED fitting to the {\it Herschel} 
160 to 500 $\mu$m data with a modified blackbody function was used 
to create a high-resolution ($18.2$\arcsec) $\rm H_{2}$ column density map with 
the method described in Appendix A of \citet{Palmeirim2013}.  
A power-law form was assumed for the dust opacity,  
$\kappa_{\lambda}=0.1\,(\lambda  /300\,\mu\rm{m})^{-\beta }\ cm^{2}\,g^{-1}$, 
with a fixed $\beta=2$ value \citep{Roy2014}.
Our column density map has twice higher resolution than the $36.3${\arcsec}-resolution map 
presented by \citet{Harvey2013}. 
When smoothed to the same 5{\arcmin} resolution, 
our high-resolution map is consistent within better than 10\% 
over more than 90\% of the pixels 
with the 
{\it Planck} $850\, \mu$m optical-depth map converted to an H$_2$ column density image 
using the above dust opacity law. 
A temperature-corrected 160 $\mu$m map was also derived by converting 
the original 160 $\mu$m map to an approximate 
column density image using the color-temperature map 
derived from the intensity ratio between 160 and 250 $\mu$m 
(see \citealt{Konyves2015}).

\subsection{SMT 10m single-dish observations of CO lines}

Maps of the X-shape Nebula in the $\rm ^{12}CO (2-1)$ and $\rm ^{13}CO (2-1)$ lines
were obtained  from the 
ESO Public Survey SAMPLING\footnote{http://sky-sampling.github.io} \citep{Wang2018}.  
These CO observations were carried out with the 10m submillimeter telescope (SMT) of the Arizona Radio 
Observatory in on-the-fly mode with Nyquist sampling, 
and cover thirteen $5\times 5${\arcmin} sub-fields arranged along the main `X' of the X-shape region 
(see Fig.~\ref{totalcoreplot}). 
We used GILDAS\footnote{https://www.iram.fr/IRAMFR/GILDAS/} 
to detect bad channels and calibrate the data.
A main beam efficiency of 0.7 was adopted to convert  
the measured antenna temperature ($T_{\rm A}$) scale to a main beam temperature ($T_{\rm mb}$) scale.
The effective resolution of the CO data is 36\arcsec, 
corresponding $\sim $0.09 pc at the distance of 500 pc.
The velocity resolution (channel width) is 0.33 km s$^{-1}$,  
the rms noise is less than 0.2 K,
and the pixel size of the final data cubes 
is 8\arcsec.

\section{Data analysis and results} \label{sec:analysis}
The {\it Herschel} 70, 160, 250, 350, 500 $\mu$m images, complemented by the 
temperature-corrected 160 $\mu$m image (at 13.5\arcsec resolution) and the H$_{2}$ 
column density image (at 18.2\arcsec resolution) were reprojected onto the same 
 3\arcsec pixel grid covering the same area of $\sim $0.7~deg$^{2}$.
\subsection{Source and filament extraction} 
\label{sec:preparation_extraction}

Before extraction of sources and filaments, all original images were 
background-subtracted and flattened using the \textsl{getimages} method 
\citep{Men2017}. This method equalizes the background and noise fluctuations 
across an entire image, making extractions more reliable (less prone to
spurious sources). The resulting flattened detection images were given as 
inputs to \textsl{getsources} \citep{Men2012}, a multi-wavelength, 
multi-scale source extraction package, which also includes the filament 
extraction algorithm \textsl{getfilaments} \citep{Men2013}. 
\textsl{getsources} and  \textsl{getfilaments} 
have only one user-defined parameter per image, the maximum size of the
sources of interest (see, e.g., Sect~3.2 in \citealt{Men2017}). 
Here, source extraction was performed\footnote{Version 2.190425 of \textsl{getsources} was used here.}
using a maximum size of four times the angular resolution at each wavelength.

Both \textsl{getsources} and  \textsl{getfilaments}  employ spatial decomposition to process and analyse the 
entire multi-wavelength data set. The decomposed images are cleaned of all
insignificant fluctuations and combined together to detect sources simultaneously 
in all wavebands. Having detected the sources, \textsl{getsources} measures 
their properties (e.g., fluxes, sizes) at each wavelength, in the original 
images. To identify the self-luminous point-like sources, a second 
extraction run, using only the 70 $\mu$m image to detect sources, is performed, 
measuring the fluxes of all detections in each waveband. 
Here, the high-resolution column density image (18.2\arcsec) was processed by \textsl{getfilaments} to detect the
filaments (their skeletons) and then to measure and catalog their properties.

For full details on how \textsl{getsources} and  \textsl{getfilaments} work, 
we refer the interested readers to the papers by 
\citet{Men2012} and \citet{Men2013,Men2017} that include also links
to the numerical codes.

                \begin{figure*}
   \centering
            \includegraphics[width=0.41 \textwidth]{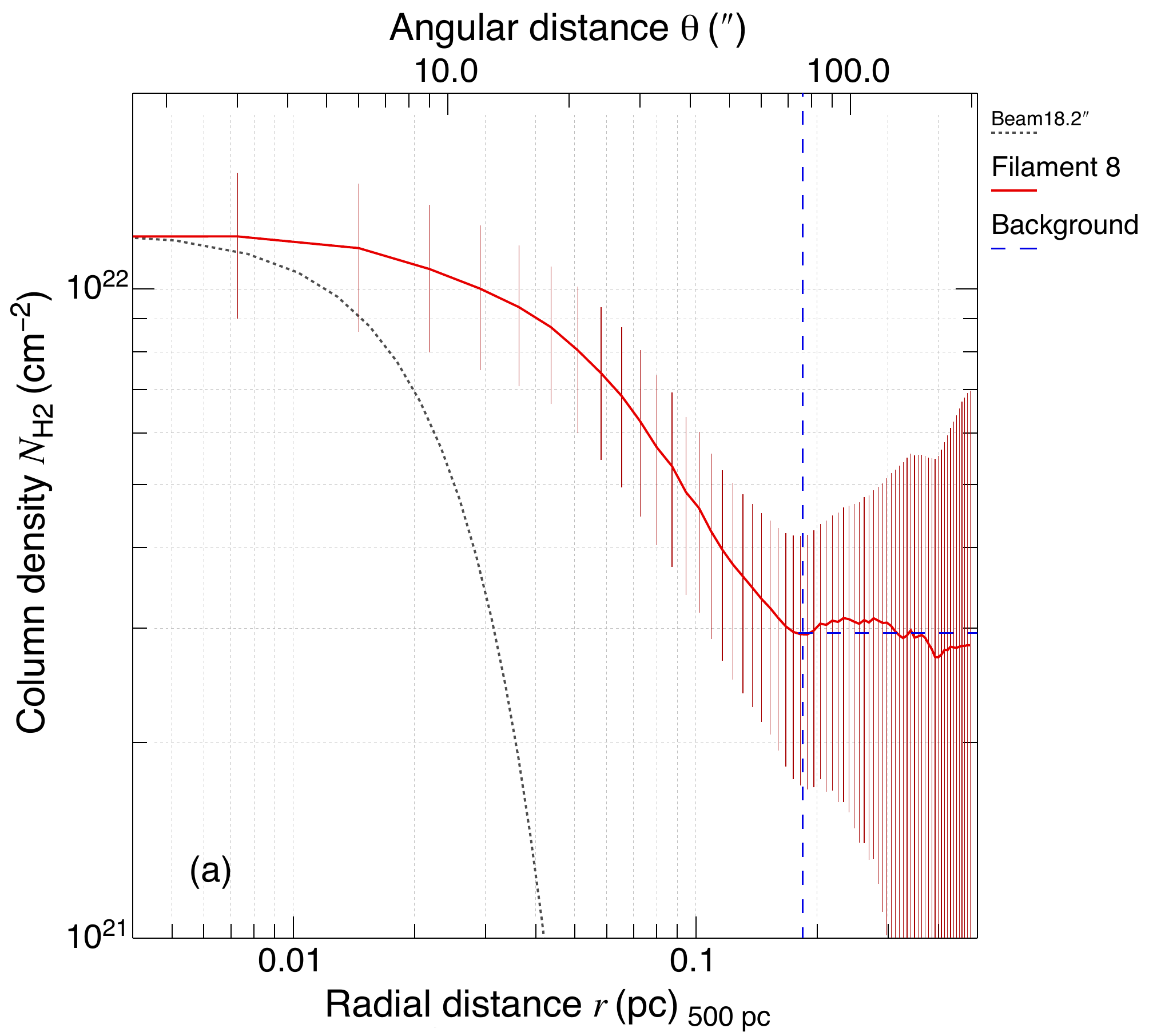}
            \hspace{1.0 cm}
            \includegraphics[width=0.41 \textwidth]{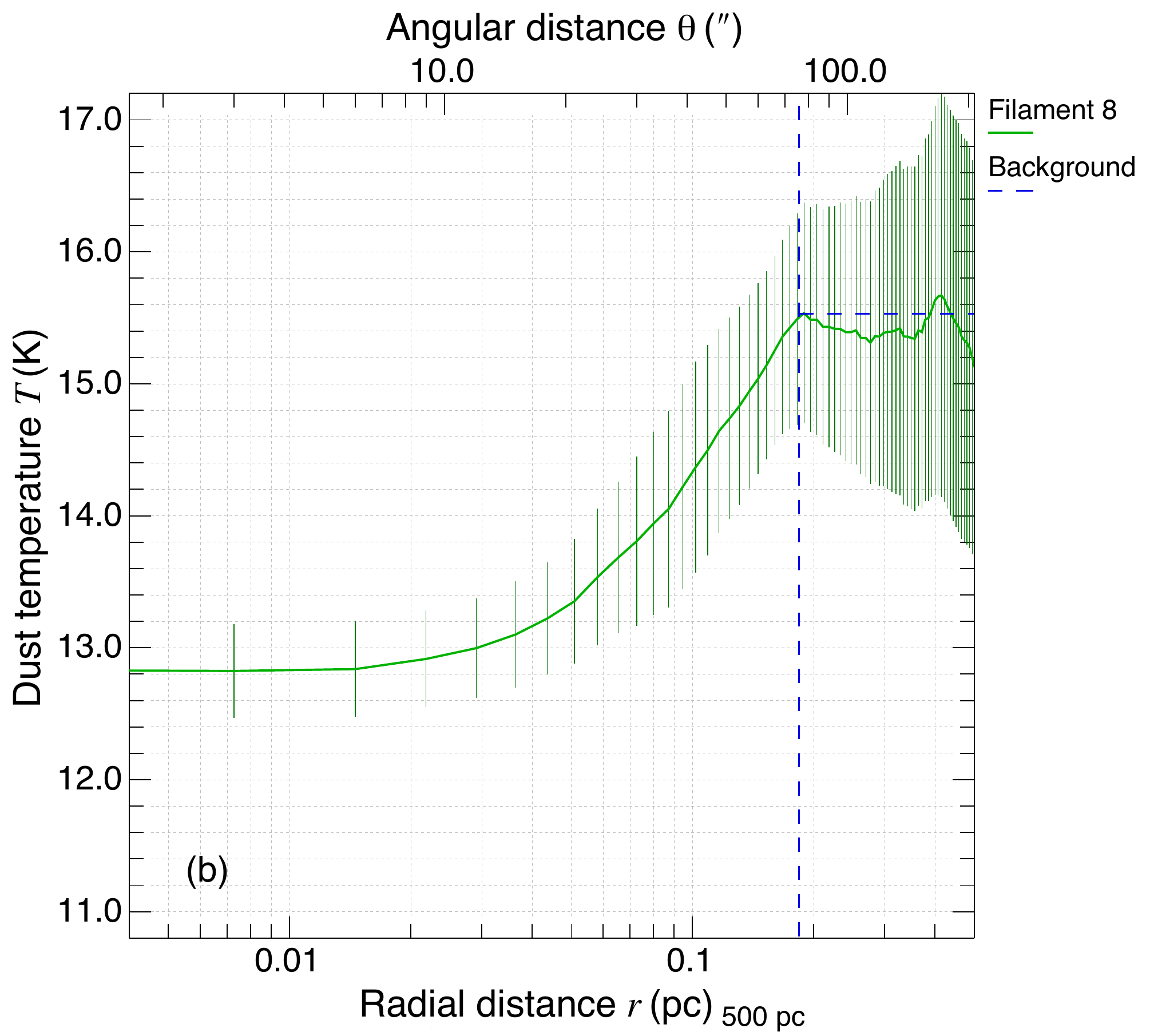}

      \caption{Median radial column density (a) and dust temperature (b) profiles of filament 8.
      Error bars correspond to the standard deviation of observed values at a given radius.
      In panel (a), the red solid line shows the median profile measured in the high-resolution (18.2\arcsec) column density map. 
      The background level (at >0.18 pc, $\sim3\times10^{21}\ \rm cm^{-2}$) is marked with a blue dashed line. 
      The median FWHM width of the filament derived from this column density profile is $0.14\pm0.03$ pc.
      In panel (b), the dust temperature measured at filament center is $\sim 13$ K, compared to 
  $\sim15.5$ K for  the local background cloud (at >0.18 pc, marked by blue dashed line).
      }
         \label{filament8}
   \end{figure*}
                  \begin{figure*}
   \centering
            \includegraphics[width=0.41 \textwidth]{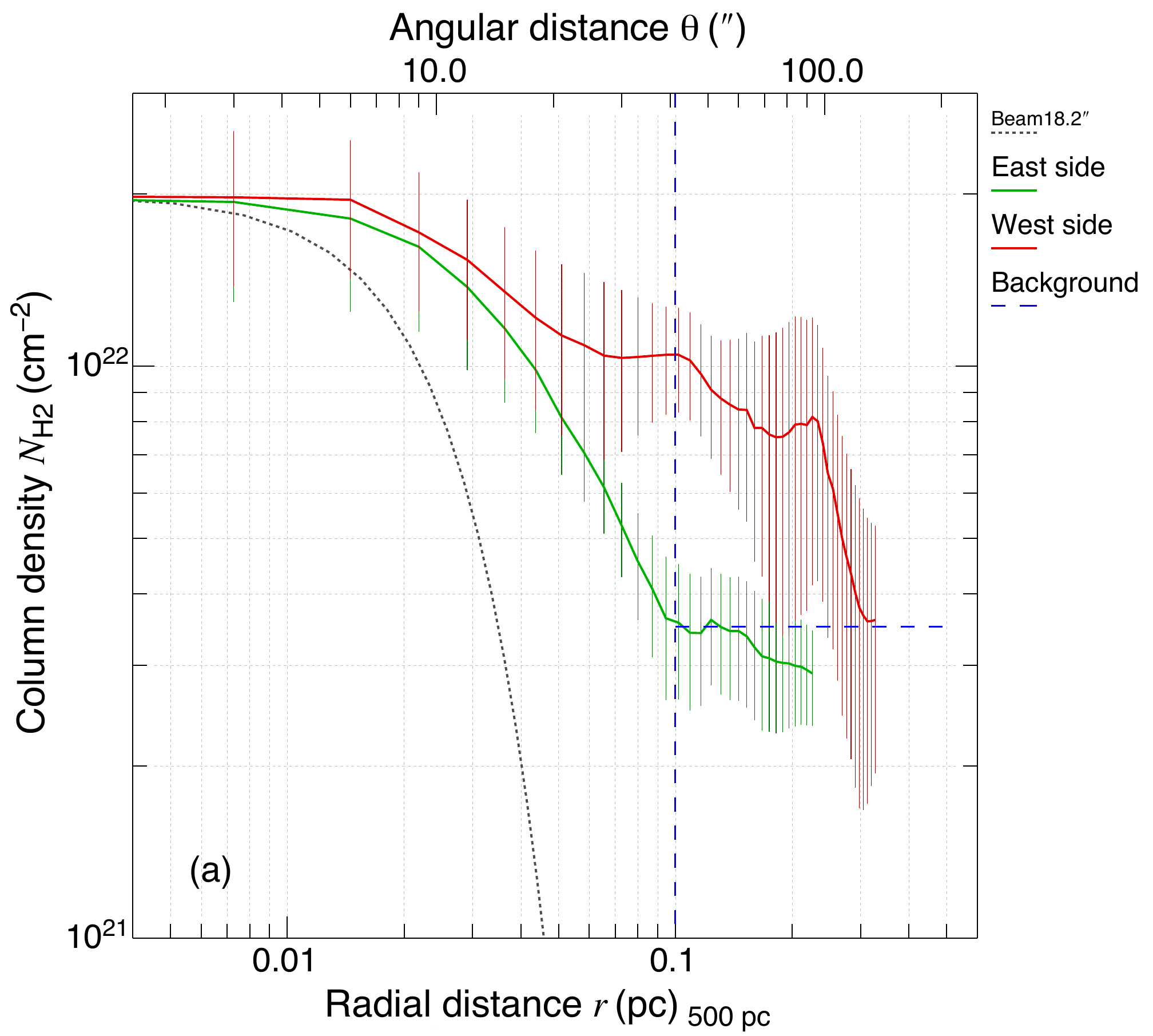}
            \hspace{1.0 cm}
            \includegraphics[width=0.41 \textwidth]{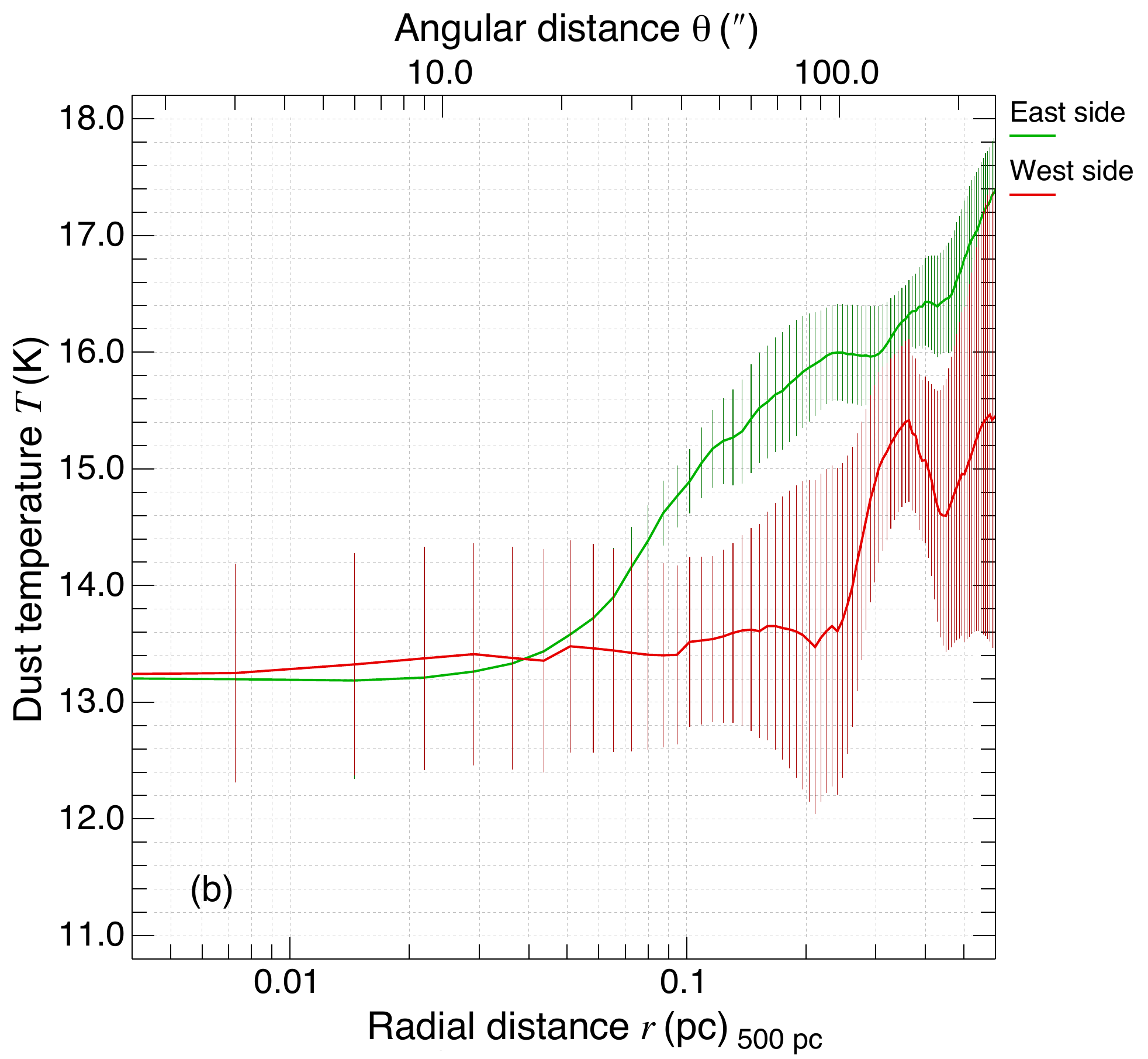}

      \caption{Median radial column density (a) and dust temperature (b) profiles measured
      on the eastern and western sides of filament 10 (green and red solid lines, respectively).
      The median FWHM width derived from the filament profile in panel (a) is $0.1\pm0.03$ pc.
      Possibly because of the effect of YSO outflows on this filament, 
      the column density profile is somewhat wider on the western side.
      The background level measured on the eastern side
      (at >0.1 pc, $\sim3.5\times10^{21}\ \rm cm^{-2}$) is marked with a blue dashed line. 
      }
         \label{filament10}
   \end{figure*}
 \begin{table*}
\small
\caption{Derived physical parameters of the ten selected filaments.}             
\label{catalogfilament}      
\centering                          
\begin{tabular}{c c c c c c c c c c c}        
\hline\hline                 
No.  &  $L_{\rm } $ & $ W_{\rm dec} $ & $AR$ &  $C$  & $T_{\rm d}^{0}$ & $N^{0}_{\rm H_{2}}$    & $n_{\rm H_{2}}$    & $M_{\rm fil}$                  & $M_{\rm line}$  \\    
      & (pc)            &  (pc)       &                &                &  (K)                   &  $\rm 10^{21}\ cm^{-2}$ & $\rm 10^{3}\ cm^{-3}$ & $(M_\sun)$ & $(M_\sun\ \rm pc^{-1})$  \\    
\hline                        
  1   &    0.53  & 0.11$\pm $0.02  & 4.4$\pm $0.8 &  0.7$\pm $0.1  & 16.4$\pm $0.2  &  0.8$\pm $0.1 &  2.2$\pm $0.4  & 1.2  &   2.3  \\
  2   &    0.48  &  0.1$\pm $0.04  & 4.4$\pm $1.6 &  0.9$\pm $0.1  & 15.9$\pm $0.3  &  1.2$\pm $0.2 &  3.5$\pm $0.6  & 1.4  &   2.9  \\
  3   &    0.8   & 0.11$\pm $0.1   & 6.8$\pm $5.7 &  0.8$\pm $0.1  & 16.8$\pm $0.2  &  0.8$\pm $0.1 &  2.3$\pm $0.6  & 1.8  &   2.2  \\
  4   &    0.64  & 0.16$\pm $0.09  &   4$\pm $2.2 &    1$\pm $0.2  & 16.3$\pm $0.4  &  1.2$\pm $0.3 &  2.4$\pm $0.8  & 2.8  &   4.3  \\
  5   &    0.93  & 0.16$\pm $0.06  & 5.8$\pm $2.3 &    2$\pm $0.3  & 15.1$\pm $0.4  &  2.4$\pm $0.3 &  4.9$\pm $1    & 8.2  &   8.8  \\
  6   &    0.95  & 0.18$\pm $0.08  & 5.1$\pm $2.2 &  1.3$\pm $0.2  &   15$\pm $0.7  &  1.8$\pm $0.3 &  3.2$\pm $0.7  & 7.2  &   7.6  \\
  7   &    1.13  & 0.13$\pm $0.03  & 8.4$\pm $1.7 &  0.9$\pm $0.2  &   16$\pm $0.5  &  1.1$\pm $0.4 &  2.7$\pm $1.1  & 3.9  &   3.5  \\
  8   &    0.98  & 0.13$\pm $0.03  & 7.1$\pm $1.7 &  5.3$\pm $0.7  & 12.9$\pm $0.4  &  9.5$\pm $1   &   22$\pm $3    & 29.1  &  29.6 \\
  9   &    0.66  &  0.1$\pm $0.04  & 5.9$\pm $2.1 &    4$\pm $1    & 13.3$\pm $0.4  &  6.8$\pm $1.5 & 22.7$\pm $8.9  & 13.2  &  19.9 \\
 10   &    0.55  & 0.09$\pm $0.03  & 5.4$\pm $1.6 &  7.6$\pm $1.1  & 13.3$\pm $0.9  &   12$\pm $1.9 & 38.5$\pm $7.7  & 14.8  &  27   \\
  
               \hline                                
\end{tabular}
\tablefoot{The filaments were detected and measured in the {\it Herschel} high-resolution ($18.2${\arcsec}) column density map.
\textbf{No.} is the index number of each filament. 
$\boldsymbol{L}$ is the length.
$\boldsymbol{W_{\rm dec}}$ is the median deconvolved FWHM width.
$\boldsymbol{AR}$ is the aspect ratio.
$\boldsymbol{C}$ is the column density contrast.
$\boldsymbol{T_{\rm d}^{0}}$ is the median centroid dust temperature.
$\boldsymbol{N^{0}_{\rm H_{2}}}$ is the median centroid $\rm H_{2}$ column density. 
$\boldsymbol{n_{\rm H_{2}}}$  is the average volume density.
$\boldsymbol{M_{\rm fil}}$ is the mass estimated by $M_{\rm fil}=M_{\rm line} \times L$. 
$\boldsymbol{M_{\rm line}}$ is the estimated line-mass. 
}   
\end{table*}
\begin{figure*}
        \centering
        \includegraphics[width=0.4 \textwidth]{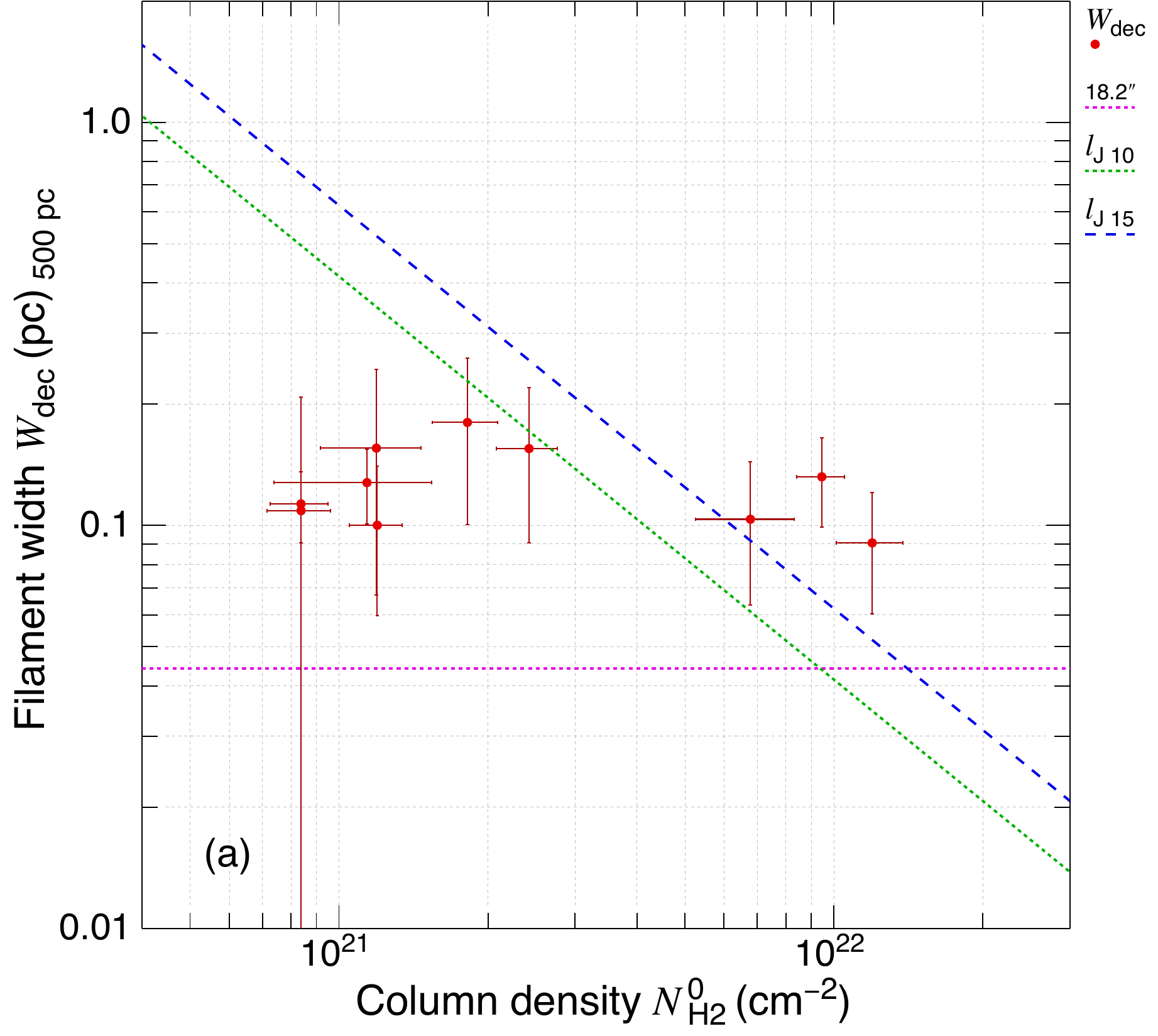}
        \hspace{0.5 cm}
        \includegraphics[width=0.42 \textwidth]{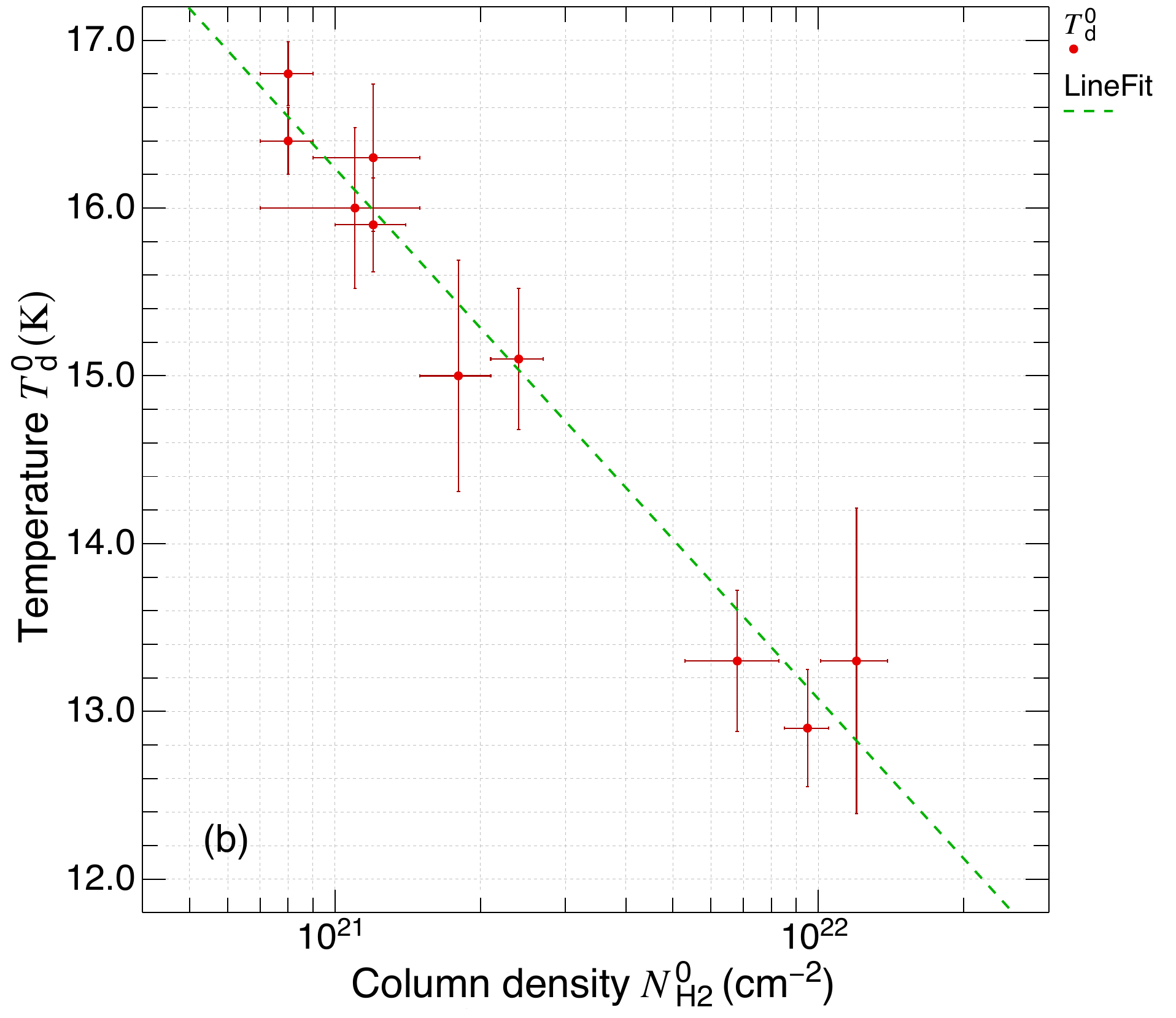}
      \caption{Plots of deconvolved FWHM width ($W_{\rm dec}$) (panel a)
      and central dust temperature ($T_{\rm d}^{0}$)  (panel b)
      against  central $\rm H_{2}$ column density ($N_{\rm H_{2}}^{0}$) 
      for the ten selected filaments of Table~\ref{catalogfilament}.
      The error bars correspond to standard deviations. 
     In panel (a), the thermal Jeans length $\lambda _{\rm J}=c_{\rm s}/\sqrt{G\rho }$ as a function of 
     column density is shown by green-dotted and blue-dashed lines for gas temperatures of 10~K and 15~K, 
     respectively, where $\rho =\mu_{\rm H_2}m_{\rm H}N_{\rm H_2}/W$ and $W\sim 0.1$ pc. 
     In panel (b), there is a clear anti-correlation between $T_{\rm d}^{0}$ and $N_{\rm H_{2}}^{0}$: 
     $T_{\rm d}^{0}\propto (-3.2\pm 0.2){\rm log}N_{\rm H_{2}}^{0}$ (green dashed line).}
         \label{widths}
\end{figure*}
\subsection{Filamentary structure selection}
\label{sec:fil_extraction}

Molecular filaments are elongated structures of gas and dust in molecular 
clouds \citep[e.g.][]{Andre2014,Andre2017}. 
We define the aspect ratios ($AR$) of a filamentary structure 
as 
\begin{equation}
AR =L/W,
\end{equation}
where $L$ is  length of the crest and $W$ is the full width at half maximum (FWHM) of the radial column density profile. 
We use the median FWHM measured along the crest in the present study. 
The column density contrast ($C$) of a filamentary structure is defined as  
\begin{equation}
C={N^{0}_{\rm H_{2}}}/{N_{\rm H_{2}^{0}}^{\rm bg}},
\end{equation}
where $N^{0}_{\rm H_{2}}$ is the median column density along the filament crest and  
${N_{\rm H_{2}}^{\rm bg}}$ is the background column density around the filament. 
$N^{0}_{\rm H_{2}}$ is measured in the source-subtracted filament map, 
while ${N_{\rm H_{2}}^{\rm bg}}$ is measured in the background map. 
Using  \textsl{getfilaments}, a network of filaments and its corresponding skeleton was 
traced in the column density map.
From this network, 
we selected 10  filamentary structures with 
aspect ratios $AR >4$ and column density contrasts $C>0.5$. 
The line mass $M_{\rm line}$ of each filament was estimated as follows:
\begin{equation} \label{mline}
M_{\rm line}\approx \mu_{\rm H_{2}} m_{\rm H} N_{\rm H_{2}}^{0} \times W,
\end{equation} 
where $\mu_{\rm H_{2}} = 2.8$ is the molecular weight per hydrogen molecule. 
$m_{\rm H}$ is the H atom mass. 
Assuming that each filament is cylindrical, we took the filament diameter 
to be the measured filament FWHM width $W$. 
 The filament average volume density was estimated as 
 \begin{equation}   \label{nh2}
 n_{\rm H_{2}}\approx N_{\rm H_{2}}^{0}/W.
\end{equation} 
We identified five subcritical filaments (filament 1--4, 7) and 
and supercritical (or transcritical) filaments (filament 5, 6, 8--10). 
Figure~\ref{filament8} shows the radial column density profile (panel a) and the temperature profile (panel b) 
of filament 8 as measured in the original column density and dust temperature maps, respectively. 
The estimated background column density toward this filament is $\rm \sim 3\times10^{21}\ cm^{2}$ 
and the centroid temperature is $\sim 13$ K.
Likewise, Fig.~\ref{filament10} shows the radial column density and temperature profiles of filament 10.
The derived physical properties of the 10 selecte filaments 
are given in Table~\ref{catalogfilament}. 
The deconvolved FWHM widths ($W_{\rm dec}$) 
of the filaments range from  
0.09 to 0.18 pc and are largely uncorrelated with 
the filament column densities ($N_{\rm H_{2}}$) (see Fig. \ref{widths}a).

The median filament width is  $W_{\rm dec}$ is $0.12\pm 0.03$ pc,
which is consistent with the common inner width $W_{\rm dec}$ $\sim 0.1$ pc 
of filaments in the {\it Herschel} Gould Belt survey (HGBS) 
 (see e.g. \citealt{Arzoumanian2011, Arzoumanian2019}). 
 The median column density contrast over the background is $C \sim 0.9\pm 0.2$ 
 for the low-density subcritical filaments and $C \sim 4\pm 0.7$  for the higher-density supercritical 
 (or transcritical) filaments. 
 Clearly, the lower contrast of subcritical filaments leads to larger measurement errors
 and larger uncertainties in their derived properties (such as their widths).
  The median line mass of subcritical filaments  is 
  $M_{\rm line}^{\rm sub} \sim 2.9\pm0.9\ M_\sun\ \rm  pc^{-1}$, 
  while that of supercritical (or transcritical) filaments is 
  $M_{\rm line}^{\rm sup} \sim 19.9\pm 10\ M_\sun\ \rm  pc^{-1}$.
  The median volume density of subcritical filaments is estimated to be 
  $n_{\rm H_{2}}^{\rm sub} \sim 2.4\pm0.6 \times 10^{3} \rm \ cm^{-3}$, 
  while that of supercritical (or transcritical) filaments is 
  $n_{\rm H_{2}}^{\rm sup} \sim 22\pm3 \times 10^{3} \rm \ cm^{-3}$.
 As the filament width $W_{\rm dec}$ is nearly uniform, 
 both $N_{\rm H_{2}}$ and $n_{\rm H_{2}}$ scale linearly with $M_{\rm line}$
 according to Eq.~(\ref{mline}) and Eq.~(\ref{nh2}). 
  The central dust temperature $T_{\rm d}$ of each filament was estimated 
  from the dust temperature map (at $36.3${\arcsec} resolution) 
  along the filament crest. 
 This dust temperature $T_{\rm d}$ is anti-correlated with $N_{\rm H_{2}}$: $T_{\rm d}\propto (-3.2\pm 0.2){\rm log}N_{\rm H_{2}}$ 
 (see Fig. \ref{widths}b).
 The median dust temperature  of subcritical filaments  is $T_{\rm d}^{\rm sub} \sim 16\,$K 
 and that of supercritical (or transcritical) filaments  3 K  lower, $T_{\rm d}^{\rm sup} \sim 13\,$K. 
For filaments including only starless cores without self-luminous protostars, 
optical depth and thus shielding against heating from the ambient radiation field 
increases, implying that high-density filaments tend to  
be colder than low-density filaments (see, e.g., \citealt{Palmeirim2013} and \citealt{Men2016}).
          \begin{figure*}
   \centering
            \includegraphics[width=0.8 \textwidth]{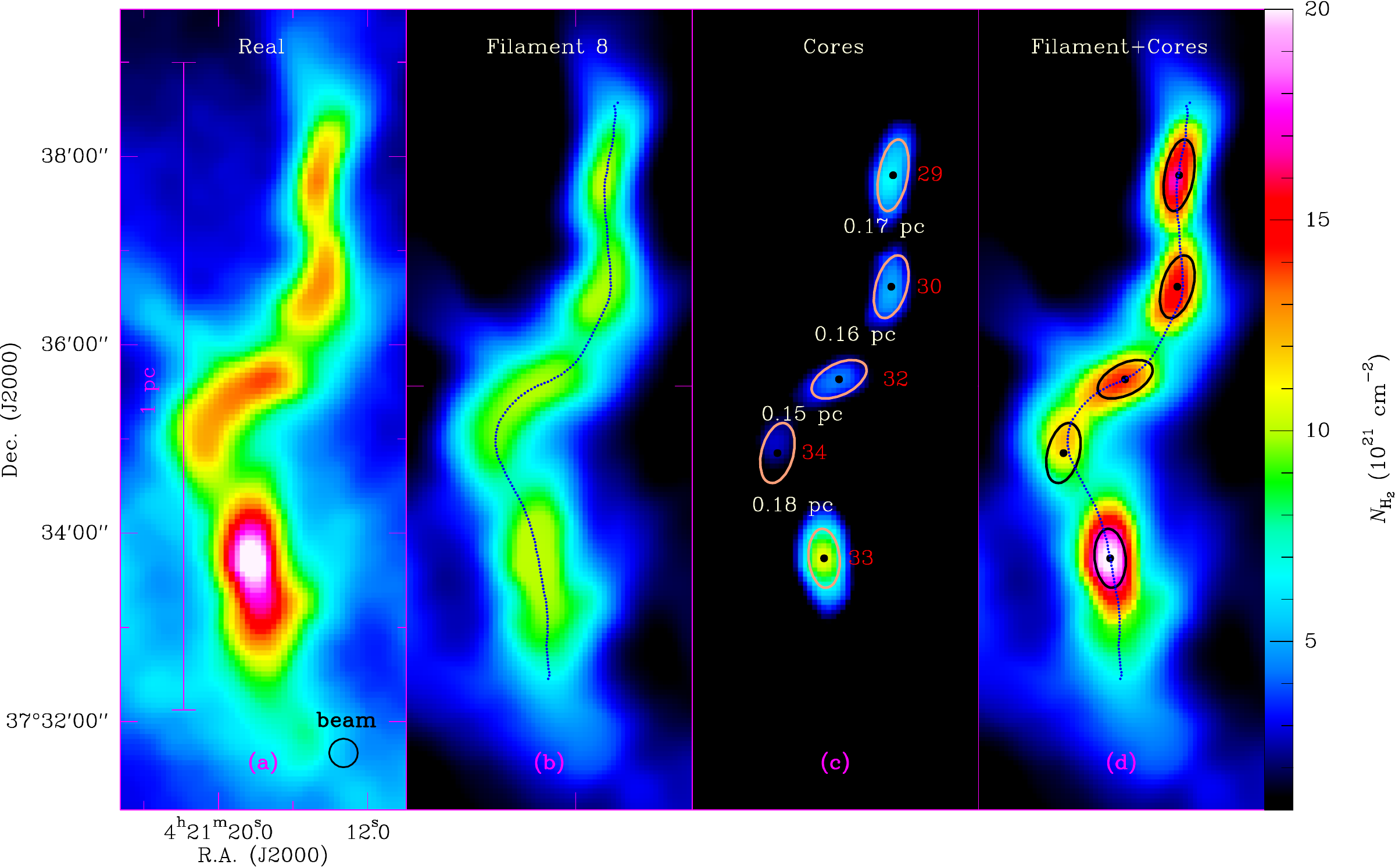}
\caption{
     High-resolution ($18.2${\arcsec}) column density views of filament 8 and its embedded cores. 
     All maps are shown from $10^{21} $ to $\rm 2 \times 10^{22}\ cm^{-2}$.
     Panel (a) shows the high-resolution ($18.2${\arcsec}) column density map. 
     Panel (b) shows a clean-background and core-subtracted filament image reconstructed over the full range of spatial scales. 
     The crest of filament is shown with a blue dotted line.
     Panel (c) shows the five robust prestellar cores identified along this filament (marked by FWHM Gaussian ellipses), 
     which are regularly spaced
     with measured projected spacings  of 0.17, 0.16, 0.15, and 0.18 pc from south to north, respectively.
     The mean core spacing is $0.17\pm0.01$ pc. 
     Panel (d) shows 
     the detected cores overlaid on the filament. 
     }
         \label{filament8-cores}
   \end{figure*}
             \begin{figure*}
   \centering
            \includegraphics[width=0.8 \textwidth]{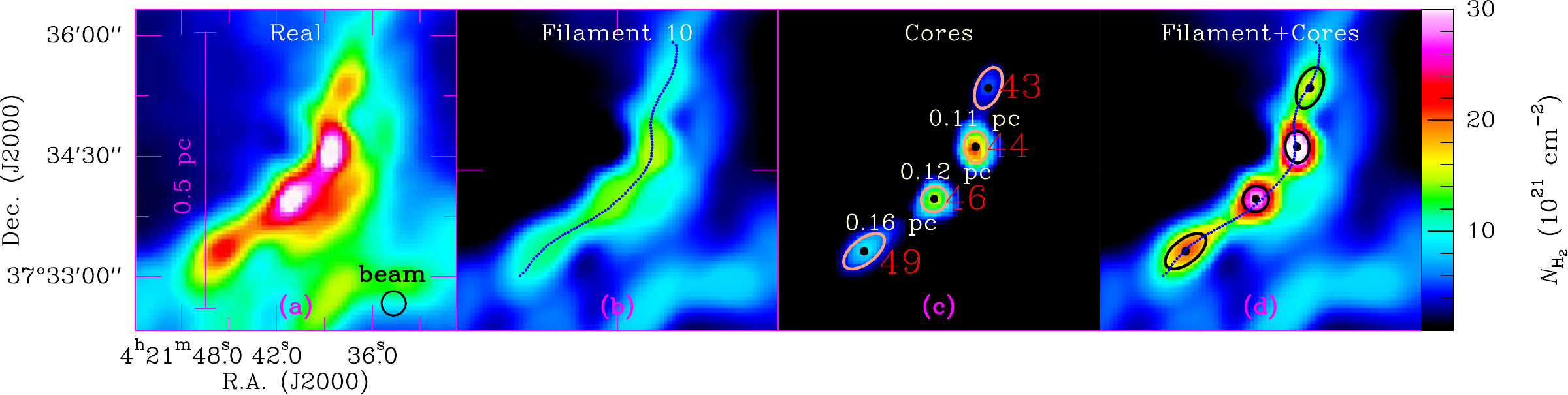}
      \caption{
      High-resolution ($18.2${\arcsec}) column density views of filament 10 and its embedded cores. 
      All maps are shown from $10^{21}$ to $\rm 3 \times 10^{22}\ cm^{-2}$.
      Two protostellar cores (core $\# $ 44, 46) and two robust prestellar cores (core $\# $ 43, 49) are detected along this filament. 
These four cores are regularly spaced with projected spacings of 0.11, 0.12, and 0.16 pc, respectively.
The average core spacing is $0.13\pm0.02$ pc. 
      }
         \label{filament10-cores}
   \end{figure*}

\begin{figure*}
   \centering
            \includegraphics[width=0.45 \textwidth]{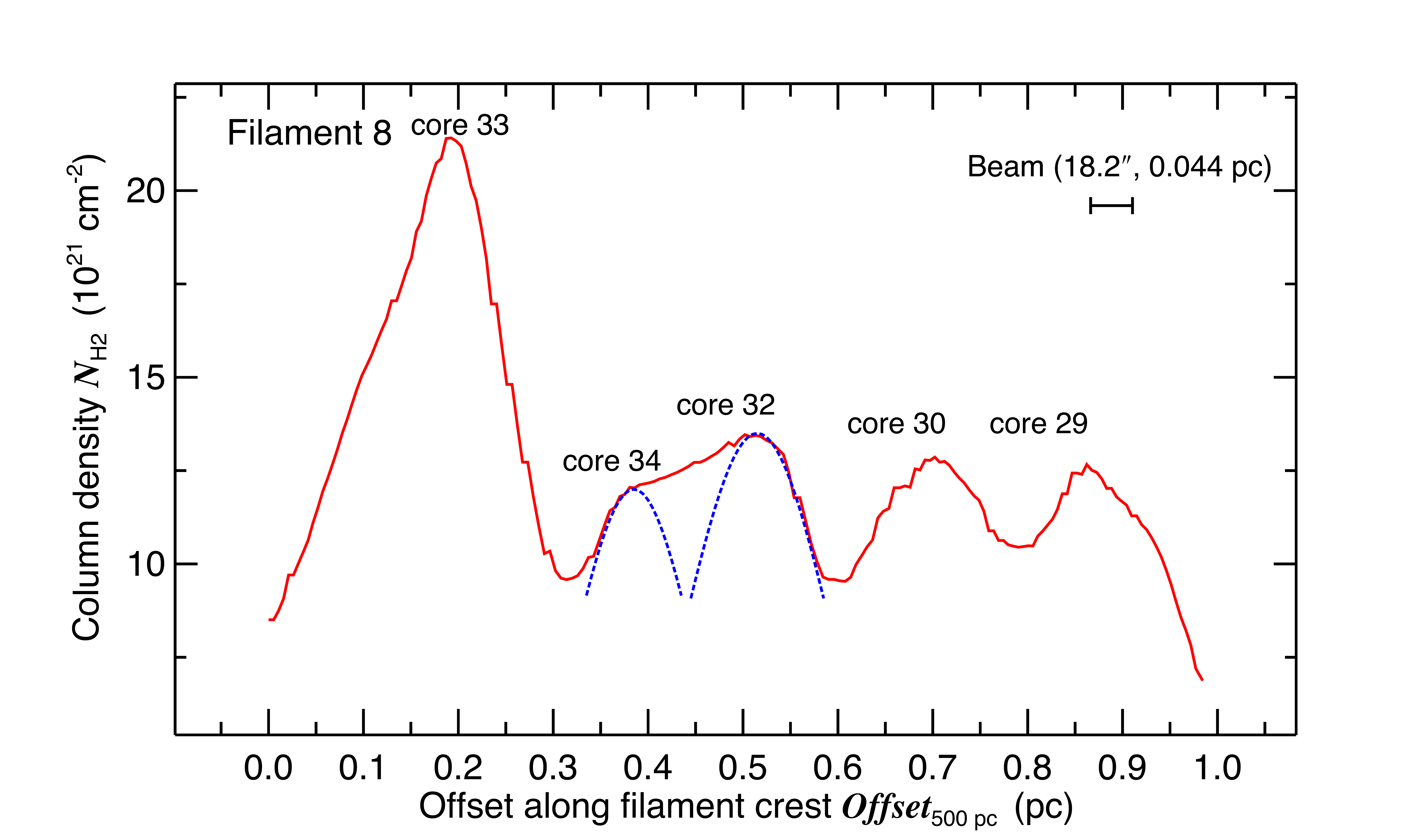}
            \includegraphics[width=0.45 \textwidth]{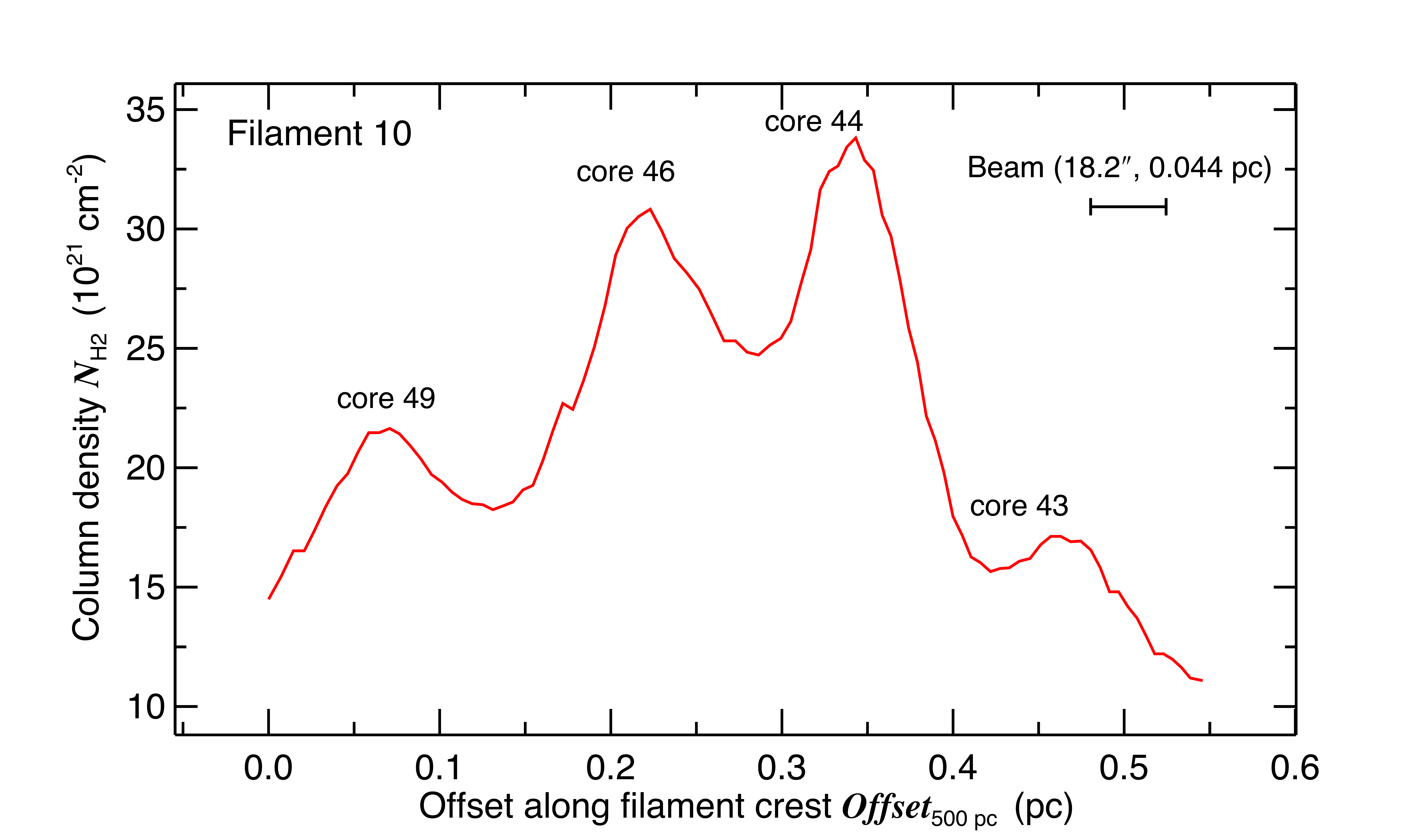}
      \caption{
      Longitudinal column density profiles of filament 8 (left panel) and filament 10  (right panel).
      In both cases, the column density profiles (red curves) were obtained from the original high-resolution (18.2\arcsec) column density map.
    In the left panel, cores $\# $ 32 and 34 partly overlap and were deblended by \textsl{getsources} 
    using an iterative method. The inferred structure of these two cores is shown by the blue dashed Gaussian profiles. 
    The four cores in the right panel are clearly separated from one another and did not require any deblending.
      }
         \label{offset}
\end{figure*}
  \begin{table*}
\small
\caption{Derived physical parameters for the cores identified along filaments 8 and 10.}             
\label{corecatalog}      
\centering                          
\begin{tabular}{c c c c c c c c c c c c }        
\hline\hline                 
 Filament & No.     & R.A.  & Dec.   &  $\textit{H}_{\rm L}$ & $\textit{H}_{\rm S}$ & PA & $R_{\rm dec}$  &  $N_{\rm H_{2}}^{\rm p}$ & $M_{\rm core}$ & $M_{\rm BE}$ & Type \\    
  &    & (J2000) & (J2000) & ({\arcsec})         &  ({\arcsec})    & ($^{\circ}$)&  (pc)    &  ($\rm 10^{21}\ cm^{-2}$) & $(M_\sun)$ & $(M_\sun)$ &\\    
\hline                        
     &29  &  04:21:14.5  &  +37:37:48  &    46.2  &    18.5  &   169.9  &    0.06  &    6.4(1)  &    0.9 & 1.1 &R-PRE\\
     &30  &  04:21:14.6  &  +37:36:37  &    41.2  &    20    &   164.1  &    0.05  &    5(1.5)  &    0.9 & 1.1 &R-PRE \\
F8   &32  &  04:21:17.4  &  +37:35:38  &    37.8  &    20    &   116.5  &    0.05  &    4.3(1.7)  &    0.7 & 1   &R-PRE \\
     &34  &  04:21:20.7  &  +37:34:51  &    39.1  &    20.5  &   165.8  &    0.05  &    2.4(1.2)  &    0.5 & 1   &R-PRE\\
     &33  &  04:21:18.2  &  +37:33:44  &    37.6  &    19.8  &     4.8  &    0.05  &   11(1.4)  &    2.3 & 1   &R-PRE\\
\hline       
      &43  &  04:21:37.7  &  +37:35:21  &    32.5  &    18.2  &   157.4  &    0.04  &    5.6(2)  &    2.8 & 0.8  & R-PRE \\
F10   &44  &  04:21:38.5  &  +37:34:37  &    24.1  &    18.2  &     3  &    0.03  &   18.7(4.3)  &    3.8 & 0.5  & PRO    \\
      &46  &  04:21:41.1  &  +37:33:58  &    20.3  &    18.7  &   130.3  &    0.02  &   15.3(4.5)  &    1.9 & 0.3 &PRO   \\
      &49  &  04:21:45.5  &  +37:33:19  &    36.6  &    18.2  &   128.2  &    0.04  &    9(3.6)  &    2.9 & 0.9   &R-PRE  \\
   \hline                                
\end{tabular}
\tablefoot{This table shows the portion of Table~\ref{totalcore} corresponding 
to the cores along filaments 8 and 10. 
\textbf{No.} is the core index number in Table~\ref{totalcore}. 
\textbf{R.A.} and \textbf{Dec.} are the centroid equatorial coordinates of the cores. 
The cores are sorted from north to south along the crest of each filament.
\bm{${H}_{\rm L}$} and \bm{${H}_{\rm S}$} are the major and minor axes 
of the elliptical Gaussian source that was fitted to each core 
in the {\it Herschel} high-resolution ($18.2${\arcsec}) column density map.
\textbf{PA} is the position angle of the major axis (measured east of north). 
$\boldsymbol{R_{\rm dec}}$ is the deconvolved radius. 
$\boldsymbol{N_{\rm H_{2}}^{\rm p}}$ is the peak column density. 
$\boldsymbol{M_{\rm core}}$ is the mass estimated from SED fitting. 
When a core is protostellar, $\boldsymbol{M_{\rm core}}$ is the protostellar envelope mass. 
$\boldsymbol{M_{\rm BE}}$ is the critical Bonnor-Ebert (BE) mass. 
\textbf{R-PRE} stands for robust prestellar core, 
\textbf{PRO} for protostellar core. }
\end{table*} 
\subsection{Selection and classification of reliable cores}\label{sec:core_selection}
Dense cores are individual cloud fragments with typical sizes $\lesssim$ 0.1 pc, 
which correspond to local overdensities and therefore local minima in the 
gravitational potential of the parent molecular cloud \citep[e.g.][]{Bergin2007, Andre2014}.
Reliable cores were selected and classified according to the detailed criteria described by \citet{Konyves2015}. 
This method has already been validated and applied to many molecular clouds 
in the {\it Herschel} Gould Belt Survey (HGBS -- \citealp{Andre2010}).
The core size is defined as the mean deconvolved FWHM diameter of an equivalent elliptical Gaussian source: 
  \begin{equation}
R_{\rm dec}=\sqrt{H_{\rm L}H_{\rm S}-O^{2}},
  \end{equation}
   where ${H}_{\rm L}$ and ${H}_{\rm S}$ are the major 
axis and minor axis of the equivalent Gaussian source, 
and  $O=18.2''$ is the angular resolution of the column density map.
The integrated flux measured for each deblended core at each wavelength by \textsl{getsources} were used 
to fit a SED with a modified blackbody function that was also used to create the column density map in Sec. \ref{sec:preparation_extraction}. 
The mass, line of sight averaged dust temperature and peak column density of each core were also estimated.
Prestellar cores are gravitationally-bound starless cores that represent fundamental units of star formation \citep{Ward2007, Bergin2007, Andre2014}. 
To first approximation, the structure of such cores resembles that of self-gravitating isothermal equilibrium 
Bonnor-Ebert (BE) spheroids \citep{Ebert1955, Bonnor1956}, bounded by the ambient pressure of the parent cloud,  
as observed in many cases \citep[e.g.][]{Alves2001,Bacmann2000, Kirk2005}. 
The BE model is useful even though real cores are not strictly isothermal \citep[e.g][]{Galli2002}
and not necessarily in hydrostatic equilibrium either \citep[e.g.][]{Ballesteros2003}.
This model has been widely used as a template to select prestellar cores, 
in, e.g., the HGBS survey \citep[see, e.g.,][]{Konyves2015} and ground-based sub-millimeter observations \citep[see, e.g.,][]{Zhang2015}.
In particular, although the critical BE mass, $M_{\rm BE}^{\rm crit}$ (maximum mass of an equilibrium isothermal sphere for 
a given temperature and ambient pressure), differs conceptually from the virial mass, it provides a good approximation 
to the latter for unmagnetized, thermal cores \citep[see, e.g.,][]{Li2013}.
The critical BE mass can be expressed as \citep{Bonnor1956}
  \begin{equation}
 M_{\rm BE}^{\rm crit} \approx 2.4\, R_{\rm BE}\, c_{\rm s}^{2}/G,
  \end{equation}
  where $R_{\rm BE}$ is the BE radius.  
Here, we used the deconvolved FWHM size measured in the column density map to estimate the core radius.   
 Assuming an ambient cloud temperature of 10 K, 
 the isothermal sound speed $c_{\rm s}$ is $\sim 0.19 \ \rm km\ s^{-1}$. 
  When the ratio $\alpha_{\rm BE}=M_{\rm BE}^{\rm crit}/M_{\rm core}\leq  2$,    
 the starless core was deemed to be self-gravitating and classified as a robust prestellar core.  
 Following \citet{Konyves2015}, an empirical size-dependent ratio 
$\alpha_{\rm BE,emp}\leq 5 \times ({O}_{N_{\rm H_2}}/{H}_{N_{\rm H_2}})^{0.4}$,
where ${O}_{N_{\rm H_2}}$ and ${H}_{N_{\rm H_2}}$ are the beam size (18.2\arcsec) 
of the high-resolution column density map and the core FWHM in this map, respectively, 
is also considered to select additional candidate prestellar cores.
A protostellar core is a dense core in which there is at least one protostar 
in the half-power column density contour.  
We identified 24 unbound starless cores, 20 robust prestellar cores, 
11 additional candidate prestellar cores, and two protostellar cores.   
The core positions are overlaid on the column density map in Fig.~\ref{totalcoreplot}  
and the physical parameters of the cores are given in Table~\ref{totalcore}. 
About 45\% of the prestellar cores lie in thermally supercritical filaments,
$\sim$ 45\% in transcritical filaments, while the rest ($\sim$ 10\%) of the prestellar cores are observed 
toward clumpy cloud structures. 
Unbound starless cores are only observed toward subcritical filamentary structures.
The derived core masses range from 0.04 to 5.8  $M_\sun$, with a mean value of 0.8 $M_\sun$. 
The deconvolved core radii  range from 0.02 to 0.12 pc, with a mean value of 0.06 pc. 
       \begin{figure*}
      \centering
      \includegraphics[width=0.32 \textwidth]{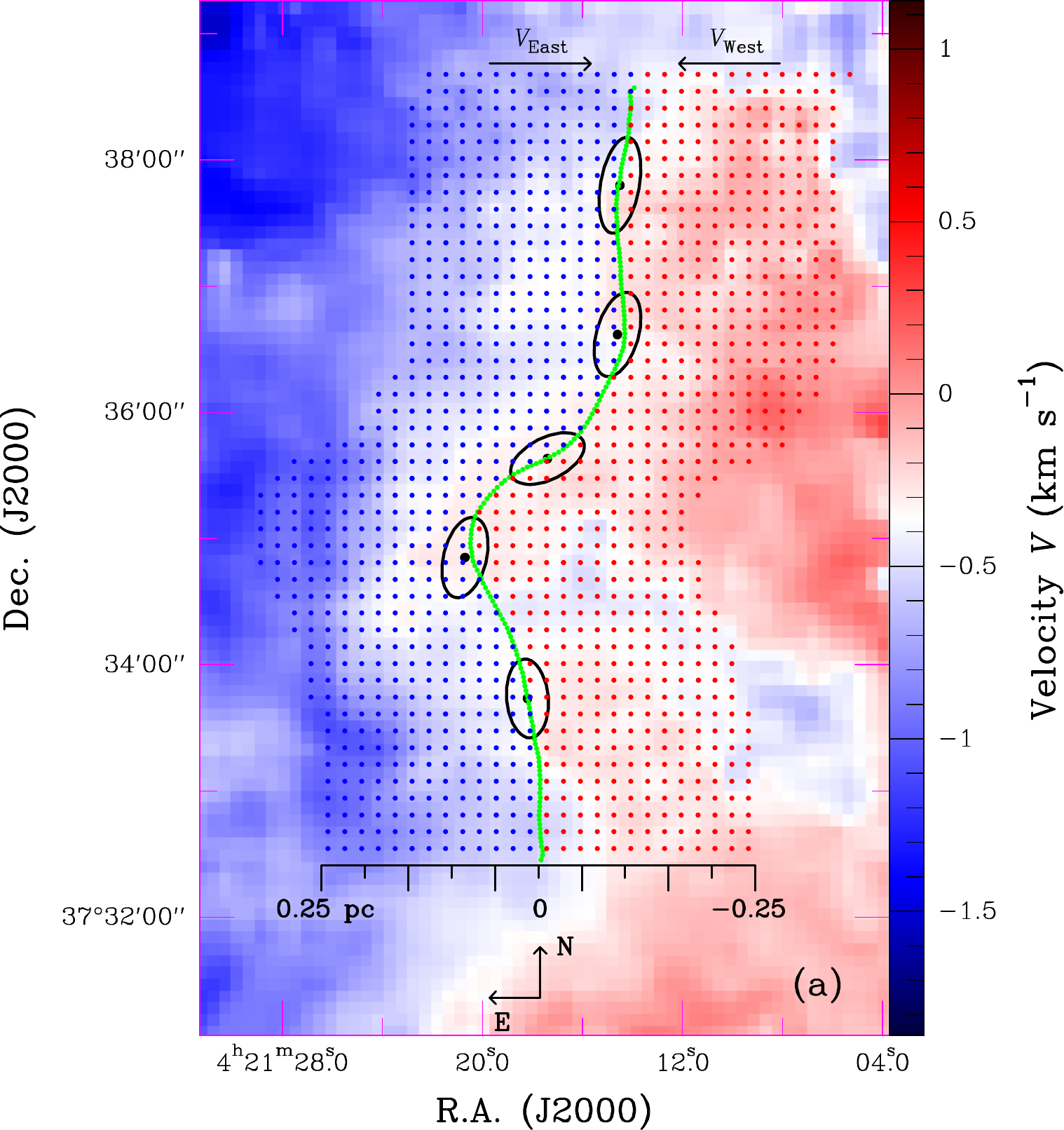}
      \includegraphics[width=0.32 \textwidth]{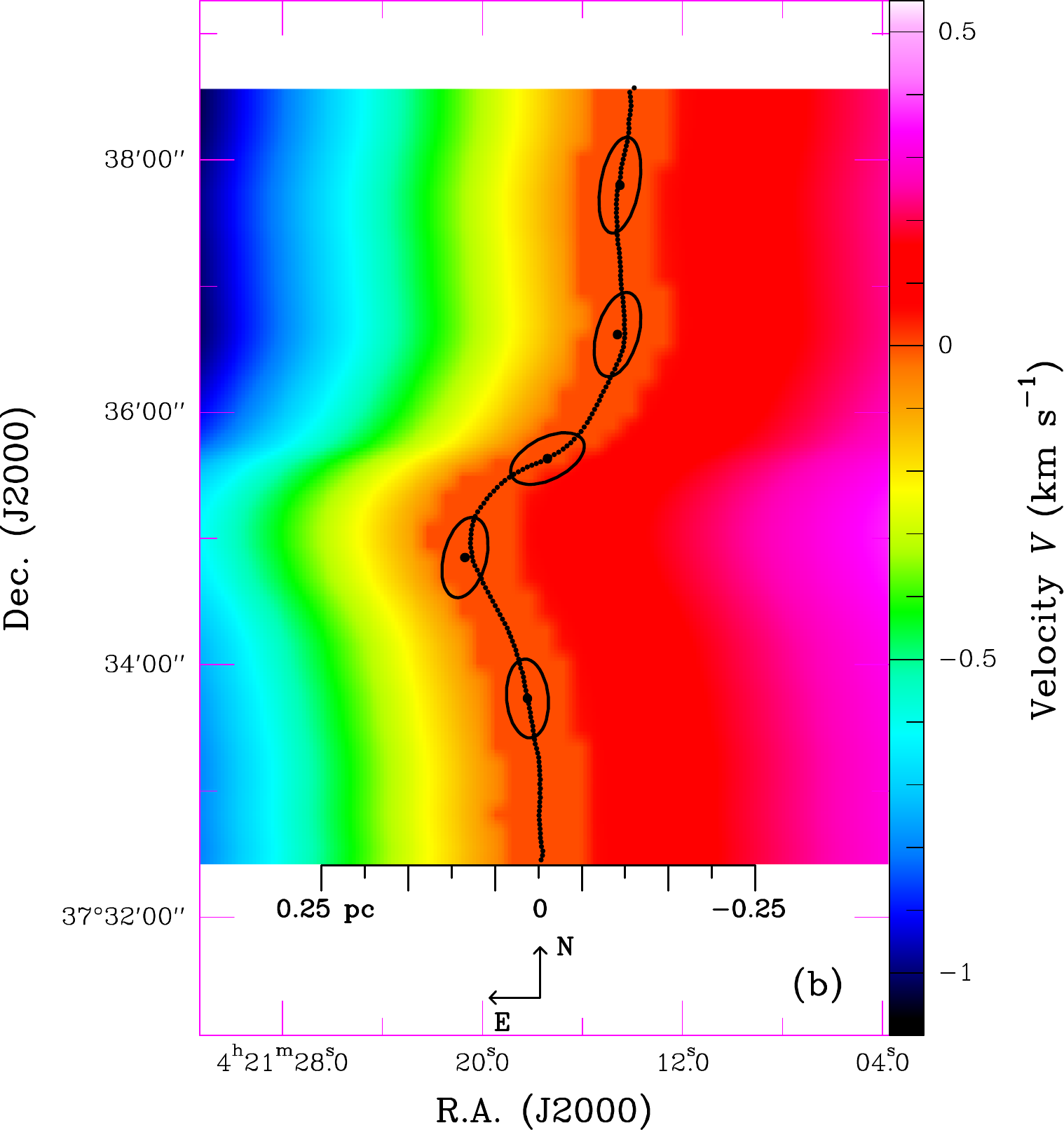}
      \includegraphics[width=0.32 \textwidth]{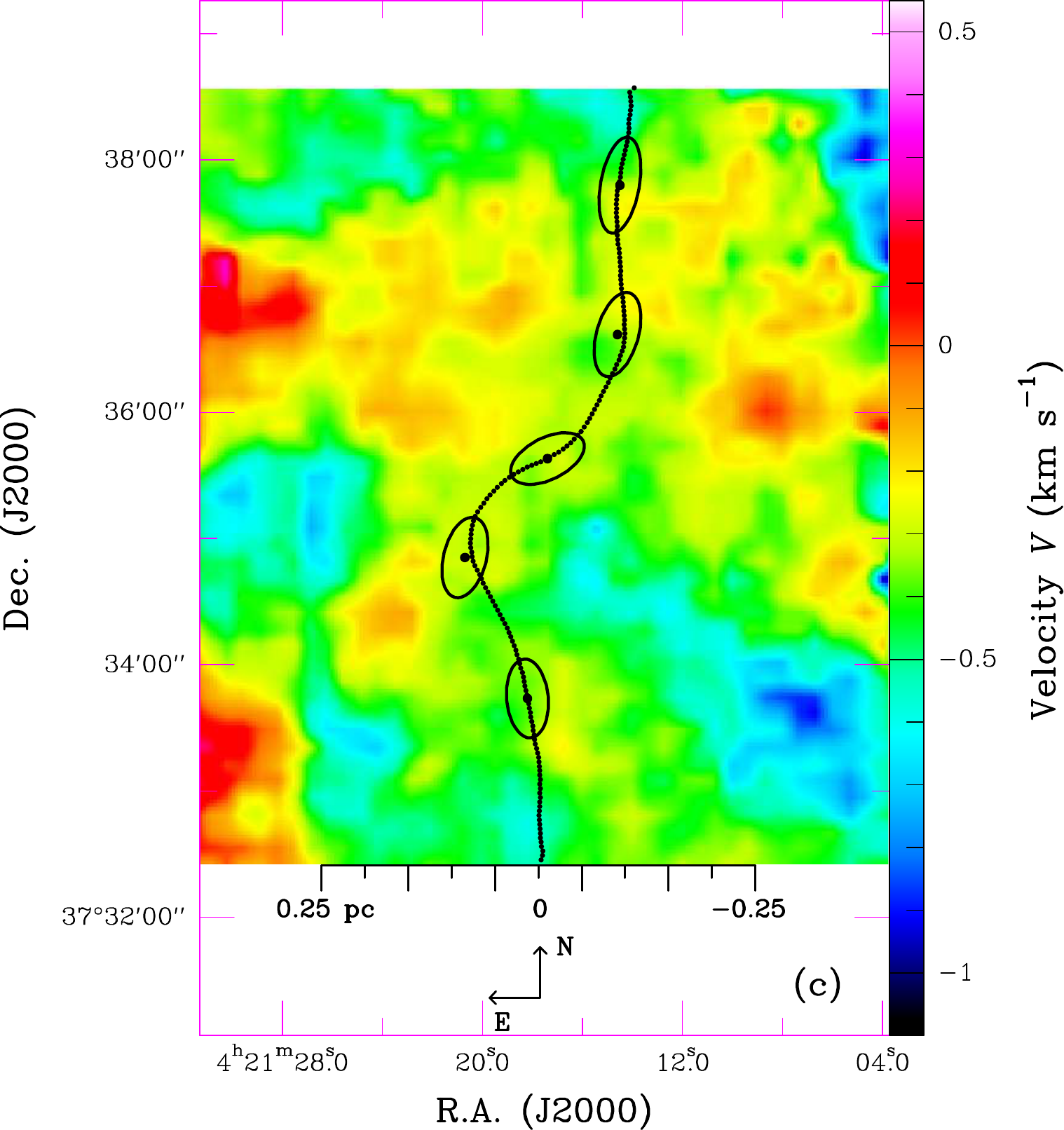}
      \caption{
            Views of the transverse velocity gradient across filament 8.
      Panel (a):   Map of centroid velocities observed in  $\rm ^{13}CO(2-1)$,  
      shown from $V_{\rm LSR} = -1.86$ $\rm km\ s^{-1}$ (blue) to $V_{\rm LSR} = +1.14$ $\rm km\ s^{-1}$ (red). 
      The crest of the filament is  marked by a green line.
      The cores identified with \textsl{getsources} are shown by black ellipses, whose 
      sizes correspond to the (major and minor) FWHM diameters of the equivalent Gaussian sources.  
      Panel (b): Map of the linear transverse-velocity gradient model fitted 
      on the blue-shifted eastern side  ($2.17\times X_{\rm offset}\rm\  km\ s^{-1}$) and 
      red-shifted western side ($0.74\times X_{\rm offset}\rm\  km\ s^{-1}$) of the filament.
      Panel (c): Map of centroid velocities obtained after subtracting the transverse velocity gradient. 
      }
         \label{f8vxmap}
   \end{figure*}
 \begin{figure}
 \centering
\includegraphics[width=0.35 \textwidth]{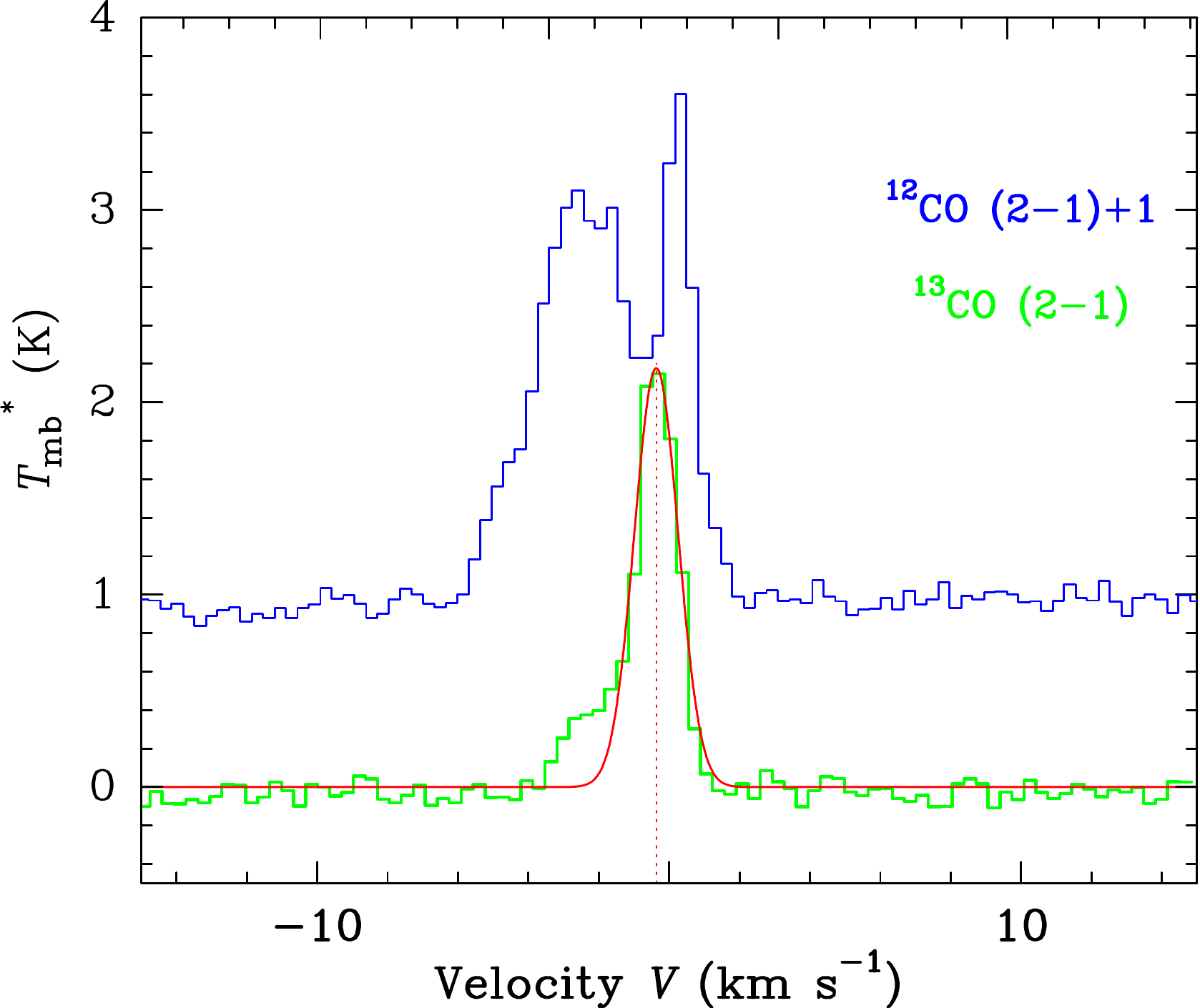}
\caption{Mean $^{12}$CO(2--1) (blue) and $^{13}$CO(2--1) (green) spectra observed with the SMT 
along the crest of  filament 8. 
A $+1$ K offset has been added to the corrected main beam temperature ($T_{\rm mb}^{*}$) 
of the $^{12}$CO(2--1) spectrum. 
The red curve shows a Gaussian fit to the $^{13}$CO(2--1) line profile.
The vertical red dashed line marks the line centroid velocity of the $^{13}$CO(2--1) spectrum ($-0.36$ km s$^{-1}$) .}
\label{f8crestsmt}
\end{figure} 
\begin{figure}
      \centering
      \vspace{-0.5 cm}
         \includegraphics[width=0.41 \textwidth]{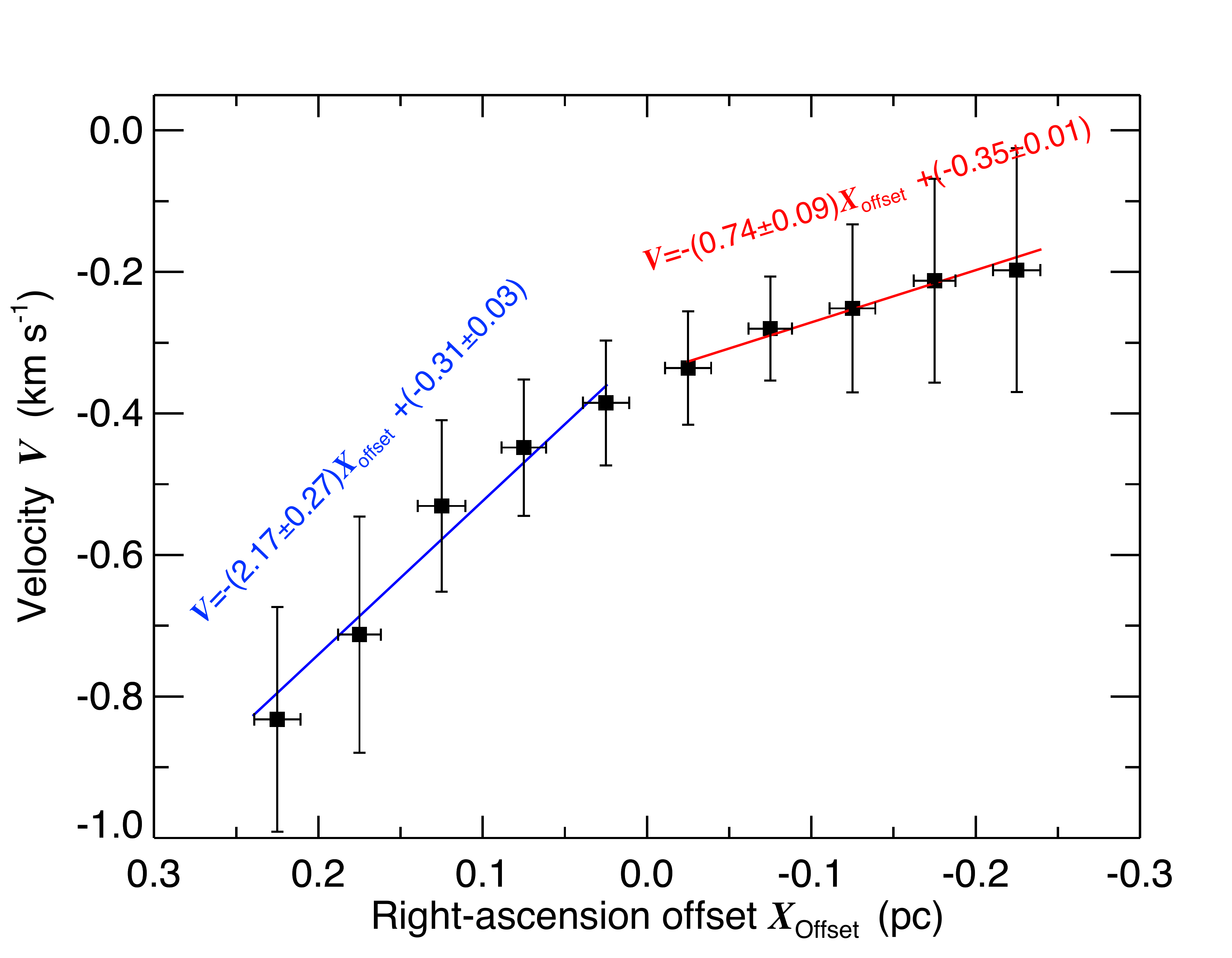}
      \caption{
     Crest-averaged centroid velocity as a function of right-ascension 
      offset $X_{\rm offset}$  from the crest of filament 8.
    This shows the man transverse velocity gradient observed across filament 8, 
   is $\sim$$ 2.17\pm0.27\rm\  km\ s^{-1}\ pc^{-1} $ on the eastern, blue-shifted side 
   and $\sim$$ 0.74\pm0.09\rm\  km\ s^{-1}\ pc^{-1} $ on the western, red-shifted side.
   }
         \label{f8vx}
   \end{figure}
A remarkable string of five regularly-spaced robust prestellar cores 
is observed along the crest of filament~8 
(see Fig.~\ref{filament8-cores}). The mean projected spacing between these five cores 
($\# $ 29, 30, 32, 34, 33) is $\sim 0.17\pm 0.1$ pc. 
(see Fig.~\ref{offset}), and their typical mass is $\sim 0.8\, M_\odot $ 
(see Table~\ref{corecatalog}). 
To test whether the quasi-periodic pattern of core spacings observed along filament~8  
may also occur for a randomly-distributed population of cores,  we used the publicly available
python code FRAGMENT \citep{Clarke2019} to construct 10\,000 random realizations 
of 5 cores randomly-placed along a 0.98-pc-long filament and to compare the resulting overall distribution 
of nearest-neighbor separations (NNS) with the observed NNS distribution using a Kolmogovov-Smirnov (K-S) test.
The results indicate that there is a very low probability, $p \sim 0.001$ (equivalent to $> 3\sigma$ in Gaussian statistics),  
that the observed quasi-periodic pattern may arise from an intrinsically random distribution of cores. 
This means that the quasi-periodic pattern of cores along filament~8 is highly significant.

Another string of four regularly-spaced cores is observed along the crest of filament 10 (see Fig.~\ref{filament10-cores}), 
including two robust prestellar cores (core $\# $ 43 and 49) and two protostellar cores (core $\# $ 44 and 46). 
Here, the mean projected core spacing is $\sim 0.13\pm0.02$ pc (see Fig.~\ref{offset}). 
Based on a K-S test similar to that described above for filament~8, the probability that the observed quasi-periodic pattern may arise 
from a random distribution of cores is $p \sim 0.008$ (equivalent to $> 2.5\sigma$ in Gaussian statistics). 
Thus, the quasi-periodic pattern of cores along filament~10 is also significant, 
although not as highly significant as in filament~8.

\subsection{Transverse velocity gradient across filament 8}\label{sec:velo-grad}
To investigate the presence of velocity gradients toward filament~8, we selected all SMT $\rm ^{13}CO (2-1)$ spectra with a signal-to-noise ratio S/N > 4.
After subtracting a baseline from each spectrum, we fitted a Gaussian profile to 
estimate the centroid velocity at each position and created a centroid velocity map (see Fig.~\ref{f8vxmap}). 
The average  $\rm ^{13}CO (2-1)$ and $\rm ^{12}CO (2-1)$ spectra observed with the SMT along 
the crest of filament~8 are shown in Fig.~\ref{f8crestsmt}.
Based on the centroid velocity map (Fig.~\ref{f8vxmap}a), we constructed an average transverse position--velocity plot across filament~8 (Fig.~\ref{f8vx}). 
To do so, we selected all $\rm ^{13}CO (2-1)$ spectra observed in a 0.5-pc-wide strip about 
the filament crest  and 
grouped them in bins of 0.05~pc 
to calculate a crest-averaged centroid velocity as a function of radial offset from the filament crest (Fig.~\ref{f8vx}).
The average centroid velocity observed along the crest of filament 8 
is $\sim -0.36\ \rm km\ s^{-1}$.
Blue-shifted gas is distributed to the east of the filament crest.
The average centroid velocity observed 0.25 pc to the east of the crest is $-0.87\ \rm \ km\ s^{-1}$,  
corresponding to a relative velocity of $\sim -0.51\ \rm km\ s^{-1}$ with respect to the filament. 
Red-shifted gas is observed to the west of the filament crest. 
The average velocity observed 0.25 pc to the west of the crest is $\sim -0.21\ \rm km\ s^{-1}$, corresponding 
to a relative velocity $\sim +0.15\ \rm km\ s^{-1}$ with respect to the filament. 
The best-fit linear velocity gradient is $\nabla V_{\rm east}=2.17\pm0.27 \rm\ km \ s^{-1}\ pc^{-1}$ on the eastern side 
and $\nabla V_{\rm west}=0.74\pm0.09 \rm \ km\ s^{-1}\ pc^{-1}$ on the western side.  
There is also a weak longitudinal velocity gradient along the crest of filament 8,  
from north to south of $\sim 0.21 \pm0.03\rm\  km\ s^{-1}\ pc^{-1}$, consistent with that reported by \citet{Imara2017} on larger scales.
We stress that the amplitude of the longitudinal velocity gradient is an order of magnitude lower 
than the amplitude of the transverse gradient, however. 
The observed longitudinal velocity gradient is also a factor of $\sim$2--4 weaker 
than that measured toward the filaments of the SDC13 infrared dark cloud, 
a hub-filament system with a morphology very similar to the X-shape region \citep{Peretto2014}. 

 \section{Discussion}\label{sec:discussion}

The {\it Herschel} data have revealed the presence of at  least two quasi-periodic chains of dense cores (along filaments 8 and 10)
in the X-shape region, with a typical projected core spacing $\sim 0.15\, $pc comparable to (or only $\sim \,$30--40\% higher than) 
the filament inner width in both cases.
 The length of filament 8 is $\sim $1 pc 
 and its line-mass is $\sim 30\ M_\sun\ \rm pc^{-1}$, roughly twice the critical line-mass 
 ($M_{\rm line,crit}\sim 16\ M_\sun\ {\rm pc^{-1}}$) for an isothermal gas cylinder at 10\, K. 
Filament 10 has about the same mass per unit length but is 
about half the length of filament 8 (see Table~\ref{catalogfilament}).
In contrast to filament 8 which includes only prestellar cores,  
filament 10 harbors two embedded protostars and may be thus be slightly more evolved, 
at a somewhat later fragmentation stage.
While filament 8 exhibits a pronounced transverse velocity gradient (Sect~\ref{sec:velo-grad} and Fig.~\ref{f8vx}), 
there is no clear evidence of such a gradient toward filament 10, but this may be due 
to confusion of the $\rm ^{12}CO(2-1)$ data by the outflows from the two embedded protostars. 
In this section, we discuss the implications of these results for our understanding of fragmentation in star-forming filaments.

 \subsection{Evidence of accretion into filament 8?}\label{sec:accretion}
Blue-shifted and red-shifted $^{12}$CO(2--1) gas is distributed to the east and to  the west of filament~8, respectively, 
a pattern observed up to at least $\pm 0.4\,$pc on either side of the filament. 
This is tracing a transverse velocity gradient of about $\rm 1.5 \ km\ s^{-1}\ pc^{-1} $ on average
(see Fig.~\ref{f8vxmap}a),  
may a priori arise from large-scale accretion, rotation, 
or shearing motions (or a combination of such motions).
While bulk rotation in most Galactic molecular clouds is known to cause velocity gradients 
$\rm \lesssim 3\ km\ s^{-1}\ pc^{-1} $ \citep[e.g.][]{Phillips1999,Braine2018}, the velocity gradient observed here
is unlikely to be due to solid-body rotation of filament 8 about its main axis for the following reasons. 
First, solid-body rotation would yield the same linear gradient on either side of the filament, which is not the case here (see Fig.~\ref{f8vx}). 
Second, the moment of inertia ${\rm I_{\Delta}}$ derived from the (column) density profile of the filament 
is very large when estimated up to $R_{\rm max} = 0.4\,$pc from the filament axis ${\Delta}$:

$${\rm I_{\Delta}}(R_{\rm max}) = \int_{0}^{R_{\rm max}} \rho(r) \, 2\pi r^3 dr \times L, $$

\noindent
where $L$ is the length of the filament.
If the observed velocity gradient is interpreted 
in terms of rotation at an angular velocity $\rm \Omega \sim 1.5 \ km\ s^{-1}\ pc^{-1}  \sim 4.9 \times 10^{-14} \ rad\ s^{-1}$, 
this would imply a very high rotational energy per unit length. 
Accordingly, the ratio of rotational ($\cal{E}_{\rm rot}$) to gravitational ($\cal{W} $) energy would also be very high for the rotating filament/cloud system: 

$$\cal{E}_{\rm rot}/\cal{W} \equiv \frac{{\rm 1/2}\, {\rm I_{\Delta}}\, {\rm \Omega}^{\rm 2}}{{\rm G\, M_{line}^2}\, {\rm L}} \approx {\rm 0.5}. $$

\noindent 
Such a high $\cal{E}_{\rm rot}/\cal{W} $ ratio is unrealistic because the centrifugal force would then severely distort the filament 
and result in a (column) density profile with secondary peaks and significantly shallower than the observed profile of Fig.~\ref{filament8}a 
(see Fig.~1 of \citealp{Recchi2014} for a normalized angular velocity $\tilde{\Omega} = \sqrt{\frac{2}{\pi G \rho_{\rm c}}}\, \Omega \sim 0.5$ 
in the case of filament 8).
Shearing motions induced by large-scale interstellar turbulence may lead to the formation of filamentary structures 
with transverse velocity gradients, although in this case the resulting filaments are expected to be non-self-gravitating 
or subcritical \citep{Hennebelle2013}, in contrast to filament~8. 
Moreover, the dimensionless parameter $C_v \equiv \frac{\Delta V^2}{G\, M_{\rm line} }$, where 
$\Delta V \sim 0.33\, \rm km\, s^{-1} $ is half the velocity difference across the filament, 
is close to 1 here ($C_v \sim 0.85$), which suggests that the velocity gradient is not due to 
to the effect of supersonic turbulence but related to the self-gravity of the filament \citep{Chen2020}. 

The most likely interpretation of the transverse velocity gradient is that it results from gravitational accretion 
motions into filament 8 within a surrounding sheet-like 
cloud structure.
A similar pattern has been reported before for other filament systems, such as 
  Taurus B211/B213 \citep{Palmeirim2013,Shimajiri2019},
  the Serpens cloud \citep{Fern2014,Dhabal2018}, or
   IRDC 18223 \citep{Beuther2015}.
 Here, the projected velocity gradient observed on the eastern side ($2.17\pm0.27 \rm\ km \ s^{-1}\  pc^{-1}$) is larger than 
 that on the western side ($0.74\pm0.09 \rm\ km \ s^{-1}\  pc^{-1}$).
 This is reminiscent of the transverse velocity gradient across the Taurus B211/B213 filament
 which is also asymmetric \citep{Palmeirim2013,Shimajiri2019}. 
 In the case of the B211/B213 filament, such a velocity gradient was successfully modeled by 
 \citet{Shimajiri2019} as arising from gravitational infall of background gas in a parent shell-like structure 
 with different inclination angles to the line of sight on either side of the filament.  
 We suggest that a similar model applies to filament 8 and can estimate the mass accretion rate 
 into the filament as follows. 
 
We define $R_{\rm out}$ to be the outer radius of the filament, outside of which the (column) density 
 profile (Fig.~\ref{filament8}a)  starts to fluctuate 
 and to be dominated by the background cloud.
From the high-resolution ($18.2${\arcsec}) $\rm H_{2}$  column density map 
and the radial profile of Fig.~\ref{filament8}a, we estimate  $R_{\rm out} \sim $ 0.18 pc 
and an average background column density $N_{\rm Background}\sim 3\times 10^{21}\rm \ cm^{-2}$. 
The relative velocity of the blue-shifted gas  
with respect to the line-of-sight velocity at filament center 
is measured to be $V_{\rm lc}=|V_{{\rm east}; R_{\rm out}}-V_{\rm c}|\approx 0.38\ \rm km\ s^{-1}$
at  $R_{\rm out}$ on the eastern side,
while the relative velocity of the red-shifted gas  is 
$V_{\rm rc}=|V_{{\rm west}; R_{\rm out}}-V_{\rm c}|\approx 0.13\ \rm km\ s^{-1}$
at  $R_{\rm out}$ on the western side.
Assuming that the ambient gas  is in free-fall motion toward the filament 
as a result of the gravitational potential of the latter, 
the free-fall velocity at $R_{\rm out}$ can be expressed as 
$V_{\rm ff}=2\sqrt{{GM_{\rm line}{\rm ln}(R_{\rm init}/R_{\rm out})}}$ \citep{Palmeirim2013},
where $R_{\rm init}$ is the radius where the gas is initially at rest.
Here, we estimate $R_{\rm init}\sim 0.4$ pc and $V_{\rm ff} \sim 0.32\ \rm km\ s^{-1}$ at $R_{\rm out}$. 
The total mass inflow rate from both sides is then expected to be
$\dot{M}_{\rm ff} =\mu_{\rm H_{2}}m_{\rm H}N_{\rm Background}V_{\rm ff}\times 2 \approx 43\ M_\sun\ \rm Myr^{-1}\ pc^{-1}$. 
The total mass accretion rate derived from the observed velocities (ignoring projection effects) is very similar: 
  $\dot{M}_{\rm obs}=\mu_{\rm H_{2}}m_{\rm H}N_{\rm Background}(V_{\rm rc}+V_{\rm lc})\approx 35\ M_\sun\ \rm  Myr^{-1}\ pc^{-1}$.
We conclude that the accretion rate into filament 8 is $\dot{M}_{\rm line}\sim 35$--$43\ M_\sun\ \rm  Myr^{-1}\ pc^{-1} $, 
corresponding to an accretion timescale,  
$M_{\rm line}/\dot{M}_{\rm line} \approx 0.7\pm0.2$ Myr. 
Interestingly, the above accretion rate is reminiscent of the value of $30 \ M_\sun\ \rm  Myr^{-1}\ pc^{-1}$ 
found for filaments in hydrodynamic simulations of molecular cloud formation by colliding flows 
and global hierarchical collapse \citep{Gomez2014,Vazquez19}. 

\subsection{Fragmentation manner of filaments 8 and 10}\label{sec:discussion:frag}
The presence of two quasi-periodic chains of dense cores in the X-shape region 
with a projected spacing similar to the width of the parent filaments is quite remarkable.
A few other examples of quasi-periodic configurations of roughly equally-spaced dense cores or clumps
along filaments have recently been reported in the literature. 
Using ALMA dust continuum emission at 870 $\mu$m, \citet{Sanchez-Monge2014} found regularly-spaced condensations 
with a mean separation of $\sim $ 0.023 pc  along the filamentary structure G35.20-0.74N. 
 \citet{Jackson2010} mapped the ``Nessie'' Nebula, a $\sim$80-pc long filamentary infrared dark cloud (IRDC), 
 in HNC (1--0) emission with the ATNF Mopra Telescope and found a characteristic clump spacing of $\sim $ 4.5 pc 
 (see also \citealp{Mattern2018}). 
 \citet{Tafalla2015} found a quasi-periodic chain of dense cores with a typical inter-core separation of $\sim 0.1$ pc in 
 the L1495/B213 filament of the Taurus MC  (see also \citealt{Bracco2017}). 
 \citet{Wang2011} found a similar separation of $\sim 0.16$ pc between dense cores along G28.34+0.06, 
 a filamentary IRDC whose line-mass is five to eight times larger than the line masses 
 of L1495/B213 and filament~8 here. 
  \citet{Shimajiri2019a} recently reported evidence of two distinct fragmentation modes in the massive NGC~6334 filament 
 ($M_{\rm line} \sim 1000\, M_\odot $/pc), with regularly-spaced clumps separated by $\sim$0.2--0.3\,pc, themselves 
 fragmented into dense cores with a typical spacing of $\sim$0.03--0.1\,pc.
While the above filaments span a wide range of length scales and line-masses 
and may not be directly comparable, it is also possible that  dense filaments may evolve and fragment 
in a qualitatively similar manner at low and high line masses \citep[cf.][]{Shimajiri2019a}, 
and that the same underlying physical mechanism is responsible for their quasi-periodic core or clump patterns.
  
Interstellar turbulence is believed to seed a whole spectrum of density fluctuations along filaments, 
and subcritical filaments are indeed observed to harbor a Kolmogorov-like spectrum of linear density 
fluctuations \citep{Roy2015}.
In the case of thermally transcritical and supercritical filaments, self-gravity can amplify some of these density fluctuations 
beyond the linear regime, leading to prestellar core formation and protostellar collapse. 
Linear fragmentation models for the growth of density perturbations along infinitely long, isothermal equilibrium  cylinders 
with $M_{\rm lin} \approx M_{\rm line,crit}$ show that density perturbations with 
wavelengths greater than $\sim 2$ times the filament diameter can grow and 
predict a characteristic core spacing of $\sim 4 \ \times$ the filament diameter \citep[e.g.][]{Inutsuka1992}, 
or $\sim 5 \ \times$ the FWHM width \citep[e.g.][]{Fischera2012}. 
Here, the $FWHM_{\rm dec}$ widths of filaments 8 and 10 are $0.13\pm0.03$ pc and $0.09\pm0.03$ pc, 
respectively, so we would expect characteristic core spacings 
$S_{\rm theory}^{\rm fil\, 8} \sim 0.65\pm 0.15$ pc and $S_{\rm theory}^{\rm fil\, 10} \sim 0.45\pm 0.15$ pc 
in the two filaments according to these models. 
For these predictions to match the observed (projected) spacings,  
$S_{\rm obs}^{\rm fil\, 8} \sim 0.17\pm0.01$ pc and $S_{\rm obs}^{\rm fil\, 10}  \sim 0.13\pm0.03$ pc, 
the two filaments would have to be seen almost ``head-on'', with a viewing angle $\alpha $ of only 
15--17$^{\circ }$ between the filament axis and the line of sight in both cases. 
Such extreme values of $\alpha $ are quite unlikely, especially for two independent filaments. 
Assuming random inclinations to the line of sight, the probability of observing a cylindrical filament 
with a viewing angle $\alpha \leq \alpha_0$ is $p = 1 - {\rm cos\,  \alpha_0}$ or $p \sim 6\% $ for $\alpha_0 = 20^{\circ }$. 
The probability of observing two filaments (such as filaments 8 and 10 here) with $\alpha \leq \alpha_0 \approx 20^{\circ} $ 
in a sample of 5 transcritical or supercritical filaments ($\# 5, 6, 8, 9, 10$ in Table~\ref{catalogfilament}) 
is only $P = \left({^5_2}\right)\, p^2\, (1-p)^3 \approx 3.6\% $ according to binomial statistics. 
The null hypothesis that the fragmentation pattern observed in filaments 8 and 10 is consistent with 
the predictions of classical cylinder fragmentation theory \citep[e.g.][]{Inutsuka1992, Inutsuka1997} 
can thus be rejected at the $> 2\sigma $ confidence level. 
The typical projected core spacing of $\sim 0.15$ pc observed toward filaments 8 and 10 
is actually in better agreement with normal or ``spherical'' Jeans-like fragmentation. 
In this case, the expected fragmentation lengthscale is the standard Jeans length, 
which may be expressed as:
\begin{equation}
\lambda _{\rm J}=c_{\rm s} \left ( \frac{\pi }{G\rho }\right )^{1/2} =0.237{\rm\ pc}\left ( \frac{T_{\rm d}^{0}}{10\rm\ K} \right )\left ( \frac{n_{\rm H_{2}}}{10^{4}\ \rm cm^{3}} \right ),
\end{equation}
where 
$ \rho =\mu_{\rm H_2} m_{\rm H}n_{\rm H_2}$ is the (central) volume density of the filament and  
$c_{\rm s}$ is the isothermal sound speed a gas temperature $T_{\rm d}^{0}$. 
Adopting $T_{\rm d}^{0} = 13$~K and the central densities given in Table~\ref{catalogfilament}, 
the Jeans length $\lambda _{\rm J}$ is estimated to be 0.11 and 0.15 pc for filament~8 and filament~10, respectively, or 0.13 pc on average. 
This is very close to the observed value of $\sim 0.15$ pc. 

Several explanations may be proposed to account for the discrepancy between the observed core spacing 
and the predictions of classical cylinder fragmentation theory. 
First, filaments 8 and 10 are clearly not perfect, straight cylinders as they exhibit rather pronounced bends  
along their lengths (cf. Figs. ~\ref{filament8-cores} and ~\ref{filament10-cores}). 
Using numerical simulations, \citet{Gritschneder2017} investigated the fragmentation properties of filaments 
including sinusoidal bends or longitudinal oscillations and showed that bending filaments 
are prone to ``geometrical fragmentation'', a process which generates cores at the turning points 
of the geometrical oscillation, separated by half the wavelength of the initial sinusoidal perturbation. 
Such a process may be partly at work here, although the geometrical deformations observed 
in filaments 8 and 10  (compared to straight filaments) are not purely sinusoidal, and some 
of the detected cores are apparently not located at clear bends along the filaments 
(cf. Figs. ~\ref{filament8-cores} and ~\ref{filament10-cores}). 

Second, although currently poorly constrained, magnetic fields may 
play a key role in controlling the detailed fragmentation manner of molecular filaments 
(see, e.g., \citealp{Andre2019} for a review). 
A magnetic field perpendicular to the filament axis tends to increase the wavelength 
of the most unstable fragmentation mode, i.e., the expected separation between cores, and 
may suppress fragmentation even for moderate field strengths \citep{Hanawa2017}.
In contrast, a longitudinal magnetic field does not suppress fragmentation 
and shortens the expected core spacing when expressed in units of the filament diameter 
\citep{Nakamura1993, Hanawa1993}. For a strong longitudinal magnetic field (e.g., a plasma $\beta \equiv \frac{\rho_c\, c_{\rm s}^2}{B_c^2/8\pi} \sim 0.1$), 
the expected core spacing can become comparable to the filament diameter (see Table~1 of \citealp{Nakamura1993}), 
as observed here in the case of  filaments 8 and 10. 
The required longitudinal field strength $\sim$$100\,\mu$G at densities $\sim$2--4$\ \times 10^4 \rm cm^{-3}$ 
in the center of filaments 8 and 10 is high, but not unrealistic in view of existing Zeeman measurements in MCs \citep{Crutcher2012}.
Observationally, the large-scale magnetic field tends to be perpendicular to thermally transcritical 
or supercritical filaments such as filaments 8 and 10 \citep{PlanckXXXII2016,PlanckXXXV2016}, 
but little is known about the geometry of the field in the {\it interior} of such filaments. 
There is a hint from {\it Planck} polarization data that the magnetic field direction may change 
in some cases from nearly perpendicular in the parent cloud to more parallel 
within the filament \citep{PlanckXXXIII2016}, but higher-resolution data would be needed 
to be truly conclusive. In particular, more work would be required to assess the magnetic field topology 
within filaments 8 and 10 and to clarify whether magnetic fields may account for the observed fragmentation lengthscale.

Third, neither filament~8 nor filament~10 is an isolated cloud structure in perfect hydrostatic equilibrium. 
As discussed in Sect.~\ref{sec:accretion}, 
the transverse velocity gradient seen across filament~8 (Sect.~\ref{sec:velo-grad} and Fig.~\ref{f8vx}) 
provides evidence that the filament is in the process of accreting gas 
from the ambient cloud \citep[cf.][]{ChenOstriker2014,Shimajiri2019}. 
As shown by \citet{Clarke2016}, the presence of accretion modifies the process of filament fragmentation. 
Indeed, \citet{Clarke2016} showed that the fastest growing mode 
for density perturbations in a nearly-critical accreting filament moves to shorter wavelengths, 
and can thus become significantly shorter than $\sim 4 \times$ the filament width, 
when the accretion rate $\dot{M}_{\rm line}^{\rm form}$ that the filament experienced during its formation  
(up to the point at which it became critical) exceeds $\Sigma_{\rm fil}\, c_{\rm s} / 2$. 
The latter essentially means that the filament formed dynamically as opposed to quasi-statically 
via subsonic assembling motions.  
In the case of filament 8,  the present accretion rate was estimated to be 
$\dot{M}_{\rm line} = 40\pm10\ M_\odot \ \rm pc^{-1} \ Myr^{-1}$
in  Sect.~\ref{sec:accretion} while  $\Sigma_{\rm fil}\, c_{\rm s} / 2$ is $\sim20\ M_\odot \ \rm pc^{-1} \ Myr^{-1}$. 
It is thus plausible that filament 8 formed dynamically on a relatively short timescale, 
which may explain the quasi-periodic spacing of its cores on a scale comparable to its width. 

\section{Conclusions}\label{sec:conclusions}
We performed a detailed study of filaments and cores in the X-shape Nebula of the California MC 
using a high-resolution (18.2\arcsec) column density map constructed from {\it Herschel} data,
along with $\rm ^{12}CO (2-1)$ and $\rm ^{13}CO (2-1)$ data from the SMT 10m telescope.
Our main findings may be summarized as follows:

   \begin{enumerate}
      \item 
      We selected 10 robust filaments with aspect ratios $AR >4$ and column density contrasts $C>0.5$
      from a skeleton network obtained with the \textsl{getfilaments} algorithm.
      The dust temperatures of the filaments are anti-correlated with their column densities ($N_{\rm H_{2}}$).  
      The deconvolved FWHM widths ($W_{\rm dec}$) of  the 10 filaments range from 0.09 to 0.18 pc 
      and are uncorrelated with their column densities ($N_{\rm H_{2}}$).
      The derived median filament width, $0.12\pm0.03$ pc, is consistent 
      with the common inner width of $\sim 0.1$ pc measured by \citet{Arzoumanian2011,Arzoumanian2019} 
      for {\it Herschel} filaments in nearby molecular clouds.
     \item  
     We identified two thermally supercritical filaments: filaments 8 and 10, 
     which both exhibit quasi-periodic chains of dense cores. 
     The typical projected core spacing is $\sim$ 0.15 pc, close or only $\sim$ 30--40\% higher than the filament inner width.
     Five prestellar cores form the chain structure of filament 8. 
     There is a transverse velocity gradient across filament 8, suggesting
     that this filament is accreting gas from a surrounding gas reservoir 
     with an accretion rate $\dot{M}\sim 40\pm10 \ M_\sun\, \rm Myr^{-1} \,  pc^{-1} $. 
     Filament 10 is $\sim $ 0.5 pc away from filament 8 and at a later fragmentation stage than filament 8.
     Two prestellar cores and two protostellar cores form the chain structure of this filament.
     \item
     We emphasize that classical cylinder fragmentation theory cannot account for the observed fragmentation 
     properties of filaments 8 and 10. 
     We suggest that three key factors may explain why the observed core spacing 
     is shorter than the standard fragmentation lengthscale of equilibrium filaments. 
     First, filaments 8 and 10 are not straight cylinder structures and feature bends along their crests
     which likely affect the fragmentation process.
     Second, if a longitudinal magnetic field of $\sim$$100\, \mu$G is present at the center of filaments 8 and 10,
     the characteristic fragmentation lengthscale may become comparable to the filament diameter as observed. 
     Third, at least in the case of filament~8, the presence of external accretion from the ambient cloud may enhance 
     initial density perturbations,  
     leading to a shorter core spacing compared to an isolated filament.
   \end{enumerate}
  
\begin{acknowledgements}
We would like to thank the anonymous referee for valuable comments 
which improved the quality of the paper.
The SMT data presented in this paper are based 
 on the ESO-ARO programme ID 196.C-0999(A). 
This work was carried out in the HGBS group of the Astrophysics Department (DAp/AIM) at CEA Paris-Saclay.
Guoyin ZHANG acknowledges support from a Chinese Government Scholarship (No. 201804910583). 
We also acknowledge support from the French national programs of CNRS/INSU on stellar and ISM physics (PNPS and PCMI). 
Ke Wang acknowledges support by the National Key Research and Development Program of China (2017YFA0402702, 2019YFA0405100), the National Science Foundation of China (11973013, 11721303), and a starting grant at the Kavli Institute for Astronomy and Astrophysics, Peking University (7101502287).
\end{acknowledgements}
\bibliographystyle{aa} 
 \bibliography{xshape.bib} 

\begin{thebibliography}{86}
\expandafter\ifx\csname natexlab\endcsname\relax\def\natexlab#1{#1}\fi

\bibitem[{{Alves} {et~al.}(2001){Alves}, {Lada}, \& {Lada}}]{Alves2001}
{Alves}, J.~F., {Lada}, C.~J., \& {Lada}, E.~A. 2001, \nat, 409, 159

\bibitem[{{Andr{\'e}}(2017)}]{Andre2017}
{Andr{\'e}}, P. 2017, Comptes Rendus Geoscience, 349, 187

\bibitem[{{Andr{\'e}} {et~al.}(2014){Andr{\'e}}, {Di Francesco},
  {Ward-Thompson}, {Inutsuka}, {Pudritz}, \& {Pineda}}]{Andre2014}
{Andr{\'e}}, P., {Di Francesco}, J., {Ward-Thompson}, D., {et~al.} 2014, in
  Protostars and Planets VI, ed. H.~{Beuther}, R.~S. {Klessen}, C.~P.
  {Dullemond}, \& T.~{Henning}, 27

\bibitem[{{Andr{\'e}} {et~al.}(2019){Andr{\'e}}, {Hughes}, {Guillet},
  {Boulanger}, {Bracco}, {Ntormousi}, {Arzoumanian}, {Maury}, {Bernard},
  {Bontemps}, {Ristorcelli}, {Girart}, {Motte}, {Tassis}, {Pantin},
  {Montmerle}, {Johnstone}, {Gabici}, {Efstathiou}, {Basu}, {B{\'e}thermin},
  {Beuther}, {Braine}, {Francesco}, {Falgarone}, {Ferri{\`e}re}, {Fletcher},
  {Galametz}, {Giard}, {Hennebelle}, {Jones}, {Kepley}, {Kwon}, {Lagache},
  {Lesaffre}, {Levrier}, {Li}, {Li}, {Mao}, {Nakagawa}, {Onaka}, {Paladino},
  {Peretto}, {Poglitsch}, {Rev{\'e}ret}, {Rodriguez}, {Sauvage}, {Soler},
  {Spinoglio}, {Tabatabaei}, {Tritsis}, {van der Tak}, {Ward-Thompson},
  {Wiesemeyer}, {Ysard}, \& {Zhang}}]{Andre2019}
{Andr{\'e}}, P., {Hughes}, A., {Guillet}, V., {et~al.} 2019, \pasa, 36, e029

\bibitem[{{Andr{\'e}} {et~al.}(2010){Andr{\'e}}, {Men'shchikov}, {Bontemps},
  {K{\"o}nyves}, {Motte}, {Schneider}, {Didelon}, {Minier}, {Saraceno},
  {Ward-Thompson}, {di Francesco}, {White}, {Molinari}, {Testi}, {Abergel},
  {Griffin}, {Henning}, {Royer}, {Mer{\'{\i}}n}, {Vavrek}, {Attard},
  {Arzoumanian}, {Wilson}, {Ade}, {Aussel}, {Baluteau}, {Benedettini},
  {Bernard}, {Blommaert}, {Cambr{\'e}sy}, {Cox}, {di Giorgio}, {Hargrave},
  {Hennemann}, {Huang}, {Kirk}, {Krause}, {Launhardt}, {Leeks}, {Le Pennec},
  {Li}, {Martin}, {Maury}, {Olofsson}, {Omont}, {Peretto}, {Pezzuto}, {Prusti},
  {Roussel}, {Russeil}, {Sauvage}, {Sibthorpe}, {Sicilia-Aguilar}, {Spinoglio},
  {Waelkens}, {Woodcraft}, \& {Zavagno}}]{Andre2010}
{Andr{\'e}}, P., {Men'shchikov}, A., {Bontemps}, S., {et~al.} 2010, \aap, 518,
  L102

\bibitem[{{Andrews} \& {Wolk}(2008)}]{Andrews2008}
{Andrews}, S.~M. \& {Wolk}, S.~J. 2008, {The LkH{\ensuremath{\alpha}} 101
  Cluster.}, Vol.~4 (Handbook of Star Forming Regions, Volume I: The Northern
  Sky ASP Monograph Publications, Edited by Bo Reipurth), 390

\bibitem[{{Arzoumanian} {et~al.}(2011){Arzoumanian}, {Andr{\'e}}, {Didelon},
  {K{\"o}nyves}, {Schneider}, {Men'shchikov}, {Sousbie}, {Zavagno}, {Bontemps},
  {di Francesco}, {Griffin}, {Hennemann}, {Hill}, {Kirk}, {Martin}, {Minier},
  {Molinari}, {Motte}, {Peretto}, {Pezzuto}, {Spinoglio}, {Ward-Thompson},
  {White}, \& {Wilson}}]{Arzoumanian2011}
{Arzoumanian}, D., {Andr{\'e}}, P., {Didelon}, P., {et~al.} 2011, \aap, 529, L6

\bibitem[{{Arzoumanian} {et~al.}(2019){Arzoumanian}, {Andr{\'e}},
  {K{\"o}nyves}, {Palmeirim}, {Roy}, {Schneider}, {Benedettini}, {Didelon}, {Di
  Francesco}, \& {Kirk}}]{Arzoumanian2019}
{Arzoumanian}, D., {Andr{\'e}}, P., {K{\"o}nyves}, V., {et~al.} 2019, \aap,
  621, A42

\bibitem[{{Bacmann} {et~al.}(2000){Bacmann}, {Andr{\'e}}, {Puget}, {Abergel},
  {Bontemps}, \& {Ward-Thompson}}]{Bacmann2000}
{Bacmann}, A., {Andr{\'e}}, P., {Puget}, J.~L., {et~al.} 2000, \aap, 361, 555

\bibitem[{{Ballesteros-Paredes} {et~al.}(2003){Ballesteros-Paredes}, {Klessen},
  \& {V{\'a}zquez-Semadeni}}]{Ballesteros2003}
{Ballesteros-Paredes}, J., {Klessen}, R.~S., \& {V{\'a}zquez-Semadeni}, E.
  2003, \apj, 592, 188

\bibitem[{{Bergin} \& {Tafalla}(2007)}]{Bergin2007}
{Bergin}, E.~A. \& {Tafalla}, M. 2007, \araa, 45, 339

\bibitem[{{Bernard} {et~al.}(2010){Bernard}, {Paradis}, {Marshall}, {Montier},
  {Lagache}, {Paladini}, {Veneziani}, {Brunt}, {Mottram}, {Martin},
  {Ristorcelli}, {Noriega-Crespo}, {Compi{\`e}gne}, {Flagey}, {Anderson},
  {Popescu}, {Tuffs}, {Reach}, {White}, {Benedettini}, {Calzoletti},
  {Digiorgio}, {Faustini}, {Juvela}, {Joblin}, {Joncas}, {Mivilles-Deschenes},
  {Olmi}, {Traficante}, {Piacentini}, {Zavagno}, \& {Molinari}}]{Bernard2010}
{Bernard}, J.~P., {Paradis}, D., {Marshall}, D.~J., {et~al.} 2010, \aap, 518,
  L88

\bibitem[{{Beuther} {et~al.}(2015){Beuther}, {Ragan}, {Johnston}, {Henning},
  {Hacar}, \& {Kainulainen}}]{Beuther2015}
{Beuther}, H., {Ragan}, S.~E., {Johnston}, K., {et~al.} 2015, \aap, 584, A67

\bibitem[{{Bonnor}(1956)}]{Bonnor1956}
{Bonnor}, W.~B. 1956, \mnras, 116, 351

\bibitem[{{Bracco} {et~al.}(2017){Bracco}, {Palmeirim}, {Andr{\'e}}, {Adam},
  {Ade}, {Bacmann}, {Beelen}, {Beno{\^\i}t}, {Bideaud}, {Billot}, {Bourrion},
  {Calvo}, {Catalano}, {Coiffard}, {Comis}, {D'Addabbo}, {D{\'e}sert},
  {Didelon}, {Doyle}, {Goupy}, {K{\"o}nyves}, {Kramer}, {Lagache}, {Leclercq},
  {Mac{\'\i}as-P{\'e}rez}, {Maury}, {Mauskopf}, {Mayet}, {Monfardini}, {Motte},
  {Pajot}, {Pascale}, {Peretto}, {Perotto}, {Pisano}, {Ponthieu},
  {Rev{\'e}ret}, {Rigby}, {Ritacco}, {Rodriguez}, {Romero}, {Roy}, {Ruppin},
  {Schuster}, {Sievers}, {Triqueneaux}, {Tucker}, \& {Zylka}}]{Bracco2017}
{Bracco}, A., {Palmeirim}, P., {Andr{\'e}}, P., {et~al.} 2017, \aap, 604, A52

\bibitem[{{Braine} {et~al.}(2018){Braine}, {Rosolowsky}, {Gratier}, {Corbelli},
  \& {Schuster}}]{Braine2018}
{Braine}, J., {Rosolowsky}, E., {Gratier}, P., {Corbelli}, E., \& {Schuster},
  K.~F. 2018, \aap, 612, A51

\bibitem[{{Broekhoven-Fiene} {et~al.}(2014){Broekhoven-Fiene}, {Matthews},
  {Harvey}, {Gutermuth}, {Huard}, {Tothill}, {Nutter}, {Bourke}, {DiFrancesco},
  \& {J{\o}rgensen}}]{Broekhoven2014}
{Broekhoven-Fiene}, H., {Matthews}, B.~C., {Harvey}, P.~M., {et~al.} 2014,
  \apj, 786, 37

\bibitem[{{Chambers} {et~al.}(2016){Chambers}, {Magnier}, {Metcalfe},
  {Flewelling}, {Huber}, {Waters}, {Denneau}, {Draper}, {Farrow}, {Finkbeiner},
  {Holmberg}, {Koppenhoefer}, {Price}, {Rest}, {Saglia}, {Schlafly}, {Smartt},
  {Sweeney}, {Wainscoat}, {Burgett}, {Chastel}, {Grav}, {Heasley}, {Hodapp},
  {Jedicke}, {Kaiser}, {Kudritzki}, {Luppino}, {Lupton}, {Monet}, {Morgan},
  {Onaka}, {Shiao}, {Stubbs}, {Tonry}, {White}, {Ba{\~n}ados}, {Bell},
  {Bender}, {Bernard}, {Boegner}, {Boffi}, {Botticella}, {Calamida},
  {Casertano}, {Chen}, {Chen}, {Cole}, {Deacon}, {Frenk}, {Fitzsimmons},
  {Gezari}, {Gibbs}, {Goessl}, {Goggia}, {Gourgue}, {Goldman}, {Grant},
  {Grebel}, {Hambly}, {Hasinger}, {Heavens}, {Heckman}, {Henderson}, {Henning},
  {Holman}, {Hopp}, {Ip}, {Isani}, {Jackson}, {Keyes}, {Koekemoer}, {Kotak},
  {Le}, {Liska}, {Long}, {Lucey}, {Liu}, {Martin}, {Masci}, {McLean}, {Mindel},
  {Misra}, {Morganson}, {Murphy}, {Obaika}, {Narayan}, {Nieto-Santisteban},
  {Norberg}, {Peacock}, {Pier}, {Postman}, {Primak}, {Rae}, {Rai}, {Riess},
  {Riffeser}, {Rix}, {R{\"o}ser}, {Russel}, {Rutz}, {Schilbach}, {Schultz},
  {Scolnic}, {Strolger}, {Szalay}, {Seitz}, {Small}, {Smith}, {Soderblom},
  {Taylor}, {Thomson}, {Taylor}, {Thakar}, {Thiel}, {Thilker}, {Unger},
  {Urata}, {Valenti}, {Wagner}, {Walder}, {Walter}, {Watters}, {Werner},
  {Wood-Vasey}, \& {Wyse}}]{Chambers2016}
{Chambers}, K.~C., {Magnier}, E.~A., {Metcalfe}, N., {et~al.} 2016, arXiv
  e-prints, arXiv:1612.05560

\bibitem[{{Chen} {et~al.}(2020){Chen}, {Mundy}, {Ostriker}, {Storm}, \&
  {Dhabal}}]{Chen2020}
{Chen}, C.-Y., {Mundy}, L.~G., {Ostriker}, E.~C., {Storm}, S., \& {Dhabal}, A.
  2020, \mnras, 494, 3675

\bibitem[{{Chen} \& {Ostriker}(2014)}]{ChenOstriker2014}
{Chen}, C.-Y. \& {Ostriker}, E.~C. 2014, \apj, 785, 69

\bibitem[{{Clarke} {et~al.}(2016){Clarke}, {Whitworth}, \&
  {Hubber}}]{Clarke2016}
{Clarke}, S.~D., {Whitworth}, A.~P., \& {Hubber}, D.~A. 2016, \mnras, 458, 319

\bibitem[{{Clarke} {et~al.}(2019){Clarke}, {Williams},
  {Ib{\'a}{\~n}ez-Mej{\'\i}a}, \& {Walch}}]{Clarke2019}
{Clarke}, S.~D., {Williams}, G.~M., {Ib{\'a}{\~n}ez-Mej{\'\i}a}, J.~C., \&
  {Walch}, S. 2019, \mnras, 484, 4024

\bibitem[{{Crutcher}(2012)}]{Crutcher2012}
{Crutcher}, R.~M. 2012, \araa, 50, 29

\bibitem[{{Dhabal} {et~al.}(2018){Dhabal}, {Mundy}, {Rizzo}, {Storm}, \&
  {Teuben}}]{Dhabal2018}
{Dhabal}, A., {Mundy}, L.~G., {Rizzo}, M.~J., {Storm}, S., \& {Teuben}, P.
  2018, \apj, 853, 169

\bibitem[{{Ebert}(1955)}]{Ebert1955}
{Ebert}, R. 1955, \zap, 37, 217

\bibitem[{{Evans}(1991)}]{Evans1991}
{Evans}, Neal~J., I. 1991, in Astronomical Society of the Pacific Conference
  Series, Vol.~20, Frontiers of Stellar Evolution, ed. D.~L. {Lambert}, 45

\bibitem[{{Federrath}(2016)}]{Federrath2016}
{Federrath}, C. 2016, Monthly Notices of the Royal Astronomical Society, 457,
  375

\bibitem[{{Fern{\'a}ndez-L{\'o}pez} {et~al.}(2014){Fern{\'a}ndez-L{\'o}pez},
  {Arce}, {Looney}, {Mundy}, {Storm}, {Teuben}, {Lee}, {Segura-Cox}, {Isella},
  {Tobin}, {Rosolowsky}, {Plunkett}, {Kwon}, {Kauffmann}, {Ostriker}, {Tassis},
  {Shirley}, \& {Pound}}]{Fern2014}
{Fern{\'a}ndez-L{\'o}pez}, M., {Arce}, H.~G., {Looney}, L., {et~al.} 2014,
  \apjl, 790, L19

\bibitem[{{Fischera} \& {Martin}(2012)}]{Fischera2012}
{Fischera}, J. \& {Martin}, P.~G. 2012, \aap, 542, A77

\bibitem[{{Gaia Collaboration} {et~al.}(2018){Gaia Collaboration}, {Brown},
  {Vallenari}, {Prusti}, {de Bruijne}, {Babusiaux}, {Bailer-Jones}, {Biermann},
  {Evans}, \& {Eyer}}]{Gaia2018}
{Gaia Collaboration}, {Brown}, A.~G.~A., {Vallenari}, A., {et~al.} 2018, \aap,
  616, A1

\bibitem[{{Galli} {et~al.}(2002){Galli}, {Walmsley}, \&
  {Gon{\c{c}}alves}}]{Galli2002}
{Galli}, D., {Walmsley}, M., \& {Gon{\c{c}}alves}, J. 2002, \aap, 394, 275

\bibitem[{{G{\'o}mez} \& {V{\'a}zquez-Semadeni}(2014)}]{Gomez2014}
{G{\'o}mez}, G.~C. \& {V{\'a}zquez-Semadeni}, E. 2014, \apj, 791, 124

\bibitem[{{Griffin} {et~al.}(2010){Griffin}, {Abergel}, {Abreu}, {Ade},
  {Andr{\'e}}, {Augueres}, {Babbedge}, {Bae}, {Baillie}, {Baluteau}, {Barlow},
  {Bendo}, {Benielli}, {Bock}, {Bonhomme}, {Brisbin}, {Brockley-Blatt},
  {Caldwell}, {Cara}, {Castro-Rodriguez}, {Cerulli}, {Chanial}, {Chen},
  {Clark}, {Clements}, {Clerc}, {Coker}, {Communal}, {Conversi}, {Cox},
  {Crumb}, {Cunningham}, {Daly}, {Davis}, {de Antoni}, {Delderfield}, {Devin},
  {di Giorgio}, {Didschuns}, {Dohlen}, {Donati}, {Dowell}, {Dowell}, {Duband},
  {Dumaye}, {Emery}, {Ferlet}, {Ferrand}, {Fontignie}, {Fox}, {Franceschini},
  {Frerking}, {Fulton}, {Garcia}, {Gastaud}, {Gear}, {Glenn}, {Goizel},
  {Griffin}, {Grundy}, {Guest}, {Guillemet}, {Hargrave}, {Harwit}, {Hastings},
  {Hatziminaoglou}, {Herman}, {Hinde}, {Hristov}, {Huang}, {Imhof}, {Isaak},
  {Israelsson}, {Ivison}, {Jennings}, {Kiernan}, {King}, {Lange}, {Latter},
  {Laurent}, {Laurent}, {Leeks}, {Lellouch}, {Levenson}, {Li}, {Li},
  {Lilienthal}, {Lim}, {Liu}, {Lu}, {Madden}, {Mainetti}, {Marliani}, {McKay},
  {Mercier}, {Molinari}, {Morris}, {Moseley}, {Mulder}, {Mur}, {Naylor},
  {Nguyen}, {O'Halloran}, {Oliver}, {Olofsson}, {Olofsson}, {Orfei}, {Page},
  {Pain}, {Panuzzo}, {Papageorgiou}, {Parks}, {Parr-Burman}, {Pearce},
  {Pearson}, {P{\'e}rez-Fournon}, {Pinsard}, {Pisano}, {Podosek}, {Pohlen},
  {Polehampton}, {Pouliquen}, {Rigopoulou}, {Rizzo}, {Roseboom}, {Roussel},
  {Rowan-Robinson}, {Rownd}, {Saraceno}, {Sauvage}, {Savage}, {Savini},
  {Sawyer}, {Scharmberg}, {Schmitt}, {Schneider}, {Schulz}, {Schwartz},
  {Shafer}, {Shupe}, {Sibthorpe}, {Sidher}, {Smith}, {Smith}, {Smith},
  {Spencer}, {Stobie}, {Sudiwala}, {Sukhatme}, {Surace}, {Stevens}, {Swinyard},
  {Trichas}, {Tourette}, {Triou}, {Tseng}, {Tucker}, {Turner}, {Vaccari},
  {Valtchanov}, {Vigroux}, {Virique}, {Voellmer}, {Walker}, {Ward}, {Waskett},
  {Weilert}, {Wesson}, {White}, {Whitehouse}, {Wilson}, {Winter}, {Woodcraft},
  {Wright}, {Xu}, {Zavagno}, {Zemcov}, {Zhang}, \& {Zonca}}]{Griffin2010}
{Griffin}, M.~J., {Abergel}, A., {Abreu}, A., {et~al.} 2010, \aap, 518, L3

\bibitem[{{Gritschneder} {et~al.}(2017){Gritschneder}, {Heigl}, \&
  {Burkert}}]{Gritschneder2017}
{Gritschneder}, M., {Heigl}, S., \& {Burkert}, A. 2017, \apj, 834, 202

\bibitem[{{Gro{\ss}schedl} {et~al.}(2019){Gro{\ss}schedl}, {Alves}, {Teixeira},
  {Bouy}, {Forbrich}, {Lada}, {Meingast}, {Hacar}, {Ascenso}, {Ackerl},
  {Hasenberger}, {K{\"o}hler}, {Kubiak}, {Larreina}, {Linhardt}, {Lombardi}, \&
  {M{\"o}ller}}]{Groschedl2019}
{Gro{\ss}schedl}, J.~E., {Alves}, J., {Teixeira}, P.~S., {et~al.} 2019, \aap,
  622, A149

\bibitem[{{Hanawa} {et~al.}(2017){Hanawa}, {Kudoh}, \& {Tomisaka}}]{Hanawa2017}
{Hanawa}, T., {Kudoh}, T., \& {Tomisaka}, K. 2017, \apj, 848, 2

\bibitem[{{Hanawa} {et~al.}(1993){Hanawa}, {Nakamura}, {Matsumoto}, {Nakano},
  {Tatematsu}, {Umemoto}, {Kameya}, {Hirano}, {Hasegawa}, {Kaifu}, \&
  {Yamamoto}}]{Hanawa1993}
{Hanawa}, T., {Nakamura}, F., {Matsumoto}, T., {et~al.} 1993, \apjl, 404, L83

\bibitem[{{Harvey} {et~al.}(2013){Harvey}, {Fallscheer}, {Ginsburg}, {Terebey},
  {Andr{\'e}}, {Bourke}, {Di Francesco}, {K{\"o}nyves}, {Matthews}, \&
  {Peterson}}]{Harvey2013}
{Harvey}, P.~M., {Fallscheer}, C., {Ginsburg}, A., {et~al.} 2013, \apj, 764,
  133

\bibitem[{{Hennebelle}(2013)}]{Hennebelle2013}
{Hennebelle}, P. 2013, \aap, 556, A153

\bibitem[{{Imara} {et~al.}(2017){Imara}, {Lada}, {Lewis}, {Bieging}, {Kong},
  {Lombardi}, \& {Alves}}]{Imara2017}
{Imara}, N., {Lada}, C., {Lewis}, J., {et~al.} 2017, \apj, 840, 119

\bibitem[{{Inutsuka} \& {Miyama}(1992)}]{Inutsuka1992}
{Inutsuka}, S.-I. \& {Miyama}, S.~M. 1992, \apj, 388, 392

\bibitem[{{Inutsuka} \& {Miyama}(1997)}]{Inutsuka1997}
{Inutsuka}, S.-i. \& {Miyama}, S.~M. 1997, \apj, 480, 681

\bibitem[{{Jackson} {et~al.}(2010){Jackson}, {Finn}, {Chambers}, {Rathborne},
  \& {Simon}}]{Jackson2010}
{Jackson}, J.~M., {Finn}, S.~C., {Chambers}, E.~T., {Rathborne}, J.~M., \&
  {Simon}, R. 2010, \apjl, 719, L185

\bibitem[{{Kainulainen} {et~al.}(2013){Kainulainen}, {Ragan}, {Henning}, \&
  {Stutz}}]{Kainulainen2013}
{Kainulainen}, J., {Ragan}, S.~E., {Henning}, T., \& {Stutz}, A. 2013, \aap,
  557, A120

\bibitem[{{Kainulainen} {et~al.}(2017){Kainulainen}, {Stutz}, {Stanke},
  {Abreu-Vicente}, {Beuther}, {Henning}, {Johnston}, \&
  {Megeath}}]{Kainulainen2017}
{Kainulainen}, J., {Stutz}, A.~M., {Stanke}, T., {et~al.} 2017, \aap, 600, A141

\bibitem[{{Kirk} {et~al.}(2013){Kirk}, {Myers}, {Bourke}, {Gutermuth},
  {Hedden}, \& {Wilson}}]{Kirk2013}
{Kirk}, H., {Myers}, P.~C., {Bourke}, T.~L., {et~al.} 2013, \apj, 766, 115

\bibitem[{{Kirk} {et~al.}(2005){Kirk}, {Ward-Thompson}, \&
  {Andr{\'e}}}]{Kirk2005}
{Kirk}, J.~M., {Ward-Thompson}, D., \& {Andr{\'e}}, P. 2005, \mnras, 360, 1506

\bibitem[{{K{\"o}nyves} {et~al.}(2020){K{\"o}nyves}, {Andr{\'e}},
  {Arzoumanian}, {Schneider}, {Men'shchikov}, {Bontemps}, {Ladjelate},
  {Didelon}, {Pezzuto}, {Benedettini}, {Bracco}, {Di Francesco}, {Goodwin},
  {Rygl}, {Shimajiri}, {Spinoglio}, {Ward-Thompson}, \& {White}}]{Konyves2020}
{K{\"o}nyves}, V., {Andr{\'e}}, P., {Arzoumanian}, D., {et~al.} 2020, \aap,
  635, A34

\bibitem[{{K{\"o}nyves} {et~al.}(2015){K{\"o}nyves}, {Andr{\'e}},
  {Men'shchikov}, {Palmeirim}, {Arzoumanian}, {Schneider}, {Roy}, {Didelon},
  {Maury}, {Shimajiri}, {Di Francesco}, {Bontemps}, {Peretto}, {Benedettini},
  {Bernard}, {Elia}, {Griffin}, {Hill}, {Kirk}, {Ladjelate}, {Marsh}, {Martin},
  {Motte}, {Nguy{\^e}n Luong}, {Pezzuto}, {Roussel}, {Rygl}, {Sadavoy},
  {Schisano}, {Spinoglio}, {Ward-Thompson}, \& {White}}]{Konyves2015}
{K{\"o}nyves}, V., {Andr{\'e}}, P., {Men'shchikov}, A., {et~al.} 2015, \aap,
  584, A91

\bibitem[{{Lada} {et~al.}(2017){Lada}, {Lewis}, {Lombardi}, \&
  {Alves}}]{Lada2017}
{Lada}, C.~J., {Lewis}, J.~A., {Lombardi}, M., \& {Alves}, J. 2017, \aap, 606,
  A100

\bibitem[{{Lada} {et~al.}(2009){Lada}, {Lombardi}, \& {Alves}}]{Lada2009}
{Lada}, C.~J., {Lombardi}, M., \& {Alves}, J.~F. 2009, \apj, 703, 52

\bibitem[{{Ladjelate} {et~al.}(2020){Ladjelate}, {Andr{\'e}}, {K{\"o}nyves},
  {Ward-Thompson}, {Men'shchikov}, {Bracco}, {Palmeirim}, {Roy}, {Shimajiri},
  {Kirk}, {Arzoumanian}, {Benedettini}, {Di Francesco}, {Fiorellino},
  {Schneider}, {Pezzuto}, {Motte}, \& {Herschel Gould Belt Survey
  Team}}]{Ladjelate2020}
{Ladjelate}, B., {Andr{\'e}}, P., {K{\"o}nyves}, V., {et~al.} 2020, \aap, 638,
  A74

\bibitem[{{Li} {et~al.}(2013){Li}, {Kauffmann}, {Zhang}, \& {Chen}}]{Li2013}
{Li}, D., {Kauffmann}, J., {Zhang}, Q., \& {Chen}, W. 2013, \apjl, 768, L5

\bibitem[{{Mattern} {et~al.}(2018){Mattern}, {Kainulainen}, {Zhang}, \&
  {Beuther}}]{Mattern2018}
{Mattern}, M., {Kainulainen}, J., {Zhang}, M., \& {Beuther}, H. 2018, \aap,
  616, A78

\bibitem[{{Men'shchikov}(2013)}]{Men2013}
{Men'shchikov}, A. 2013, \aap, 560, A63

\bibitem[{{Men'shchikov}(2016)}]{Men2016}
{Men'shchikov}, A. 2016, \aap, 593, A71

\bibitem[{{Men'shchikov}(2017)}]{Men2017}
{Men'shchikov}, A. 2017, \aap, 607, A64

\bibitem[{{Men'shchikov} {et~al.}(2010){Men'shchikov}, {Andr{\'e}}, {Didelon},
  {K{\"o}nyves}, {Schneider}, {Motte}, {Bontemps}, {Arzoumanian}, {Attard},
  {Abergel}, {Baluteau}, {Bernard}, {Cambr{\'e}sy}, {Cox}, {di Francesco}, {di
  Giorgio}, {Griffin}, {Hargrave}, {Huang}, {Kirk}, {Li}, {Martin}, {Minier},
  {Miville-Desch{\^e}nes}, {Molinari}, {Olofsson}, {Pezzuto}, {Roussel},
  {Russeil}, {Saraceno}, {Sauvage}, {Sibthorpe}, {Spinoglio}, {Testi},
  {Ward-Thompson}, {White}, {Wilson}, {Woodcraft}, \& {Zavagno}}]{Men2010}
{Men'shchikov}, A., {Andr{\'e}}, P., {Didelon}, P., {et~al.} 2010, \aap, 518,
  L103

\bibitem[{{Men'shchikov} {et~al.}(2012){Men'shchikov}, {Andr{\'e}}, {Didelon},
  {Motte}, {Hennemann}, \& {Schneider}}]{Men2012}
{Men'shchikov}, A., {Andr{\'e}}, P., {Didelon}, P., {et~al.} 2012, \aap, 542,
  A81

\bibitem[{{Nakamura} {et~al.}(1993){Nakamura}, {Hanawa}, \&
  {Nakano}}]{Nakamura1993}
{Nakamura}, F., {Hanawa}, T., \& {Nakano}, T. 1993, \pasj, 45, 551

\bibitem[{{Ostriker}(1964)}]{Ostriker1964}
{Ostriker}, J. 1964, The Astrophysical Journal, 140, 1056

\bibitem[{{Palmeirim} {et~al.}(2013){Palmeirim}, {Andr{\'e}}, {Kirk},
  {Ward-Thompson}, {Arzoumanian}, {K{\"o}nyves}, {Didelon}, {Schneider},
  {Benedettini}, {Bontemps}, {Di Francesco}, {Elia}, {Griffin}, {Hennemann},
  {Hill}, {Martin}, {Men'shchikov}, {Molinari}, {Motte}, {Nguyen Luong},
  {Nutter}, {Peretto}, {Pezzuto}, {Roy}, {Rygl}, {Spinoglio}, \&
  {White}}]{Palmeirim2013}
{Palmeirim}, P., {Andr{\'e}}, P., {Kirk}, J., {et~al.} 2013, \aap, 550, A38

\bibitem[{{Peretto} {et~al.}(2014){Peretto}, {Fuller}, {Andr{\'e}},
  {Arzoumanian}, {Rivilla}, {Bardeau}, {Duarte Puertas}, {Guzman Fernandez},
  {Lenfestey}, {Li}, {Olguin}, {R{\"o}ck}, {de Villiers}, \&
  {Williams}}]{Peretto2014}
{Peretto}, N., {Fuller}, G.~A., {Andr{\'e}}, P., {et~al.} 2014, \aap, 561, A83

\bibitem[{{Phillips}(1999)}]{Phillips1999}
{Phillips}, J.~P. 1999, \aaps, 134, 241

\bibitem[{{Planck Collab. Int. XXXII}(2016)}]{PlanckXXXII2016}
{Planck Collab. Int. XXXII}. 2016, \aap, 586, A135

\bibitem[{{Planck Collab. Int. XXXIII}(2016)}]{PlanckXXXIII2016}
{Planck Collab. Int. XXXIII}. 2016, \aap, 586, A136

\bibitem[{{Planck Collab. Int. XXXV}(2016)}]{PlanckXXXV2016}
{Planck Collab. Int. XXXV}. 2016, \aap, 586, A138

\bibitem[{{Poglitsch} {et~al.}(2010){Poglitsch}, {Waelkens}, {Geis},
  {Feuchtgruber}, {Vandenbussche}, {Rodriguez}, {Krause}, {Renotte}, {van
  Hoof}, {Saraceno}, {Cepa}, {Kerschbaum}, {Agn{\`e}se}, {Ali}, {Altieri},
  {Andreani}, {Augueres}, {Balog}, {Barl}, {Bauer}, {Belbachir}, {Benedettini},
  {Billot}, {Boulade}, {Bischof}, {Blommaert}, {Callut}, {Cara}, {Cerulli},
  {Cesarsky}, {Contursi}, {Creten}, {De Meester}, {Doublier}, {Doumayrou},
  {Duband}, {Exter}, {Genzel}, {Gillis}, {Gr{\"o}zinger}, {Henning},
  {Herreros}, {Huygen}, {Inguscio}, {Jakob}, {Jamar}, {Jean}, {de Jong},
  {Katterloher}, {Kiss}, {Klaas}, {Lemke}, {Lutz}, {Madden}, {Marquet},
  {Martignac}, {Mazy}, {Merken}, {Montfort}, {Morbidelli}, {M{\"u}ller},
  {Nielbock}, {Okumura}, {Orfei}, {Ottensamer}, {Pezzuto}, {Popesso},
  {Putzeys}, {Regibo}, {Reveret}, {Royer}, {Sauvage}, {Schreiber}, {Stegmaier},
  {Schmitt}, {Schubert}, {Sturm}, {Thiel}, {Tofani}, {Vavrek}, {Wetzstein},
  {Wieprecht}, \& {Wiezorrek}}]{Poglitsch2010}
{Poglitsch}, A., {Waelkens}, C., {Geis}, N., {et~al.} 2010, \aap, 518, L2

\bibitem[{{Pudritz} \& {Kevlahan}(2013)}]{Pudritz2013}
{Pudritz}, R.~E. \& {Kevlahan}, N.~K.~R. 2013, Philosophical Transactions of
  the Royal Society of London Series A, 371, 20120248

\bibitem[{{Recchi} {et~al.}(2014){Recchi}, {Hacar}, \&
  {Palestini}}]{Recchi2014}
{Recchi}, S., {Hacar}, A., \& {Palestini}, A. 2014, \mnras, 444, 1775

\bibitem[{{Roy} {et~al.}(2015){Roy}, {Andr{\'e}}, {Arzoumanian}, {Peretto},
  {Palmeirim}, {K{\"o}nyves}, {Schneider}, {Benedettini}, {Di Francesco},
  {Elia}, {Hill}, {Ladjelate}, {Louvet}, {Motte}, {Pezzuto}, {Schisano},
  {Shimajiri}, {Spinoglio}, {Ward-Thompson}, \& {White}}]{Roy2015}
{Roy}, A., {Andr{\'e}}, P., {Arzoumanian}, D., {et~al.} 2015, \aap, 584, A111

\bibitem[{{Roy} {et~al.}(2014){Roy}, {Andr{\'e}}, {Palmeirim}, {Attard},
  {K{\"o}nyves}, {Schneider}, {Peretto}, {Men'shchikov}, {Ward-Thompson}, \&
  {Kirk}}]{Roy2014}
{Roy}, A., {Andr{\'e}}, P., {Palmeirim}, P., {et~al.} 2014, \aap, 562, A138

\bibitem[{{S{\'a}nchez-Monge} {et~al.}(2014){S{\'a}nchez-Monge}, {Beltr{\'a}n},
  {Cesaroni}, {Etoka}, {Galli}, {Kumar}, {Moscadelli}, {Stanke}, {van der Tak},
  {Vig}, {Walmsley}, {Wang}, {Zinnecker}, {Elia}, {Molinari}, \&
  {Schisano}}]{Sanchez-Monge2014}
{S{\'a}nchez-Monge}, {\'A}., {Beltr{\'a}n}, M.~T., {Cesaroni}, R., {et~al.}
  2014, \aap, 569, A11

\bibitem[{{Schlafly} {et~al.}(2014){Schlafly}, {Green}, {Finkbeiner}, {Rix},
  {Bell}, {Burgett}, {Chambers}, {Draper}, {Hodapp}, {Kaiser}, {Magnier},
  {Martin}, {Metcalfe}, {Price}, \& {Tonry}}]{Schlafly2014}
{Schlafly}, E.~F., {Green}, G., {Finkbeiner}, D.~P., {et~al.} 2014, \apj, 786,
  29

\bibitem[{{Shimajiri} {et~al.}(2019{\natexlab{a}}){Shimajiri}, {Andr{\'e}},
  {Ntormousi}, {Men'shchikov}, {Arzoumanian}, \& {Palmeirim}}]{Shimajiri2019a}
{Shimajiri}, Y., {Andr{\'e}}, P., {Ntormousi}, E., {et~al.} 2019{\natexlab{a}},
  \aap, 632, A83

\bibitem[{{Shimajiri} {et~al.}(2019{\natexlab{b}}){Shimajiri}, {Andr{\'e}},
  {Palmeirim}, {Arzoumanian}, {Bracco}, {K{\"o}nyves}, {Ntormousi}, \&
  {Ladjelate}}]{Shimajiri2019}
{Shimajiri}, Y., {Andr{\'e}}, P., {Palmeirim}, P., {et~al.} 2019{\natexlab{b}},
  \aap, 623, A16

\bibitem[{{Tafalla} \& {Hacar}(2015)}]{Tafalla2015}
{Tafalla}, M. \& {Hacar}, A. 2015, \aap, 574, A104

\bibitem[{{V{\'a}zquez-Semadeni} {et~al.}(2019){V{\'a}zquez-Semadeni}, {Palau},
  {Ballesteros-Paredes}, {G{\'o}mez}, \& {Zamora-Avil{\'e}s}}]{Vazquez19}
{V{\'a}zquez-Semadeni}, E., {Palau}, A., {Ballesteros-Paredes}, J.,
  {G{\'o}mez}, G.~C., \& {Zamora-Avil{\'e}s}, M. 2019, \mnras, 490, 3061

\bibitem[{{Wang} {et~al.}(2015){Wang}, {Testi}, {Ginsburg}, {Walmsley},
  {Molinari}, \& {Schisano}}]{Wang2015}
{Wang}, K., {Testi}, L., {Ginsburg}, A., {et~al.} 2015, \mnras, 450, 4043

\bibitem[{{Wang} {et~al.}(2018){Wang}, {Zahorecz}, {Cunningham}, {T{\'o}th},
  {Liu}, {Lu}, {Wang}, {Cosentino}, {Sung}, {Sokolov}, {Wang}, {Wang}, {Zhang},
  {Li}, {Kim}, {Tatematsu}, {Testi}, {Wu}, {Yang}, \& {SAMPLING
  Collaboration}}]{Wang2018}
{Wang}, K., {Zahorecz}, S., {Cunningham}, M.~R., {et~al.} 2018, Research Notes
  of the American Astronomical Society, 2, 2

\bibitem[{{Wang} {et~al.}(2011){Wang}, {Zhang}, {Wu}, \& {Zhang}}]{Wang2011}
{Wang}, K., {Zhang}, Q., {Wu}, Y., \& {Zhang}, H. 2011, \apj, 735, 64

\bibitem[{{Ward-Thompson} {et~al.}(2007){Ward-Thompson}, {Andr{\'e}},
  {Crutcher}, {Johnstone}, {Onishi}, \& {Wilson}}]{Ward2007}
{Ward-Thompson}, D., {Andr{\'e}}, P., {Crutcher}, R., {et~al.} 2007, in
  Protostars and Planets V, ed. B.~{Reipurth}, D.~{Jewitt}, \& K.~{Keil}, 33

\bibitem[{{Yan} {et~al.}(2019){Yan}, {Zhang}, {Xu}, {Guo}, {Macquart}, {Tang},
  \& {Walsh}}]{Yan2019}
{Yan}, Q.-Z., {Zhang}, B., {Xu}, Y., {et~al.} 2019, \aap, 624, A6

\bibitem[{{Zhang} {et~al.}(2015){Zhang}, {Li}, {Hyde}, {Qian}, {Lyu}, \&
  {Wu}}]{Zhang2015}
{Zhang}, G., {Li}, D., {Hyde}, A.~K., {et~al.} 2015, Science China Physics,
  Mechanics, and Astronomy, 58, 5561

\bibitem[{{Zhang} {et~al.}(2018){Zhang}, {Xu}, {Vasyunin}, {Semenov}, {Wang},
  {Dib}, {Liu}, {Liu}, {Zhang}, {Liu}, {Wang}, {Li}, {Wu}, {Yuan}, {Li}, \&
  {Gao}}]{Zhang2018}
{Zhang}, G.-Y., {Xu}, J.-L., {Vasyunin}, A.~I., {et~al.} 2018, \aap, 620, A163

\bibitem[{{Zucker} {et~al.}(2019){Zucker}, {Speagle}, {Schlafly}, {Green},
  {Finkbeiner}, {Goodman}, \& {Alves}}]{Zucker2019}
{Zucker}, C., {Speagle}, J.~S., {Schlafly}, E.~F., {et~al.} 2019, \apj, 879,
  125

\end{thebibliography}
 \balance
 %

\begin{appendix} 
\onecolumn

\section{Core location distribution}
 \begin{figure*}
 \centering
\includegraphics[width=0.8 \textwidth]{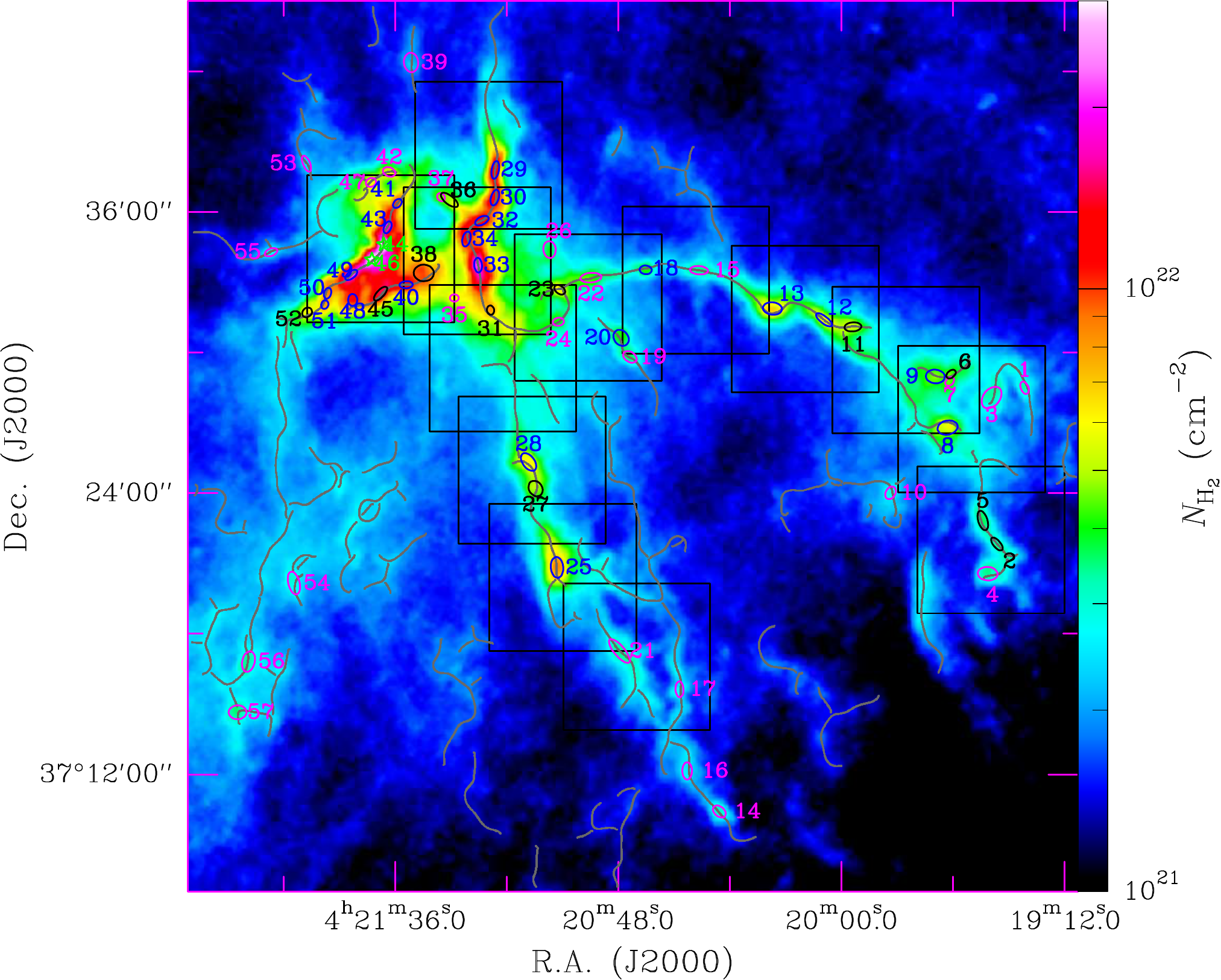}
\caption{ Positions of the 57 dense cores (FWHM Gaussian ellipses)
identified in the X-shape Nebula region 
overlaid on the  {\it Herschel} high-resolution ($18.2${\arcsec}) column density map.
Magenta ellipses mark the 24 unbound starless cores,
blue ellipses the 11 candidate prestellar cores,
black ellipses the 20 robust prestellar cores, 
and green stars the 2 protostellar cores (see Sect.~\ref{sec:core_selection}).
The skeleton of the filament network extracted in Sect.~\ref{sec:fil_extraction} is shown by grey curves. 
The thirteen ${5}'\times {5}'$ fields covered by the SMT CO observations are marked as black boxes.}
         \label{totalcoreplot}
\end{figure*} 
\begin{table*}
\small
\caption{Physical parameters of the cores identified with {\it Herschel} in the X-shape Nebula region.}             
\label{totalcore}      
\centering                          
\begin{tabular}{c c c c c c c c c c c l}        
\hline\hline                 
No.     & RA  & Dec   &  $\textit{H}_{\rm L}$ & $\textit{H}_{\rm S}$ & PA & $R_{\rm dec}$  &  $N^{\rm p}_{\rm H_{2}}$   & $N^{\rm p}_{\rm H_{2}}/N^{\rm bg}_{\rm H_{2}}$ & $M_{\rm core}$   & $M_{\rm BE}$ & Type   \\    
     & (J2000) & (J2000) & $('')$         &  $('')$    & ($^{\circ}$)&  (pc)   &  ($\rm 10^{21}\ cm^{-2}$) & & $(M_{\odot })$  & $(M_{\odot })$&            \\    
\hline                        
1   &     04:19:20.6   &  +37:28:30  &    33.5  &    19.6  &    22.6  &    0.04  &     0.9  &     0.5  &    0.09  &     0.9  &  U-STA  \\
2   &     04:19:26.7   &  +37:21:50  &    40.9  &    18.2  &    40.4  &    0.05  &     0.7  &     0.3  &     0.3  &     1    &  C-PRE  \\
3   &     04:19:27.7   &  +37:28:05  &    62.9  &    42.1  &   140.2  &    0.12  &     0.7  &     0.4  &     0.2  &     2.3  &  U-STA  \\
4   &     04:19:28.7   &  +37:20:34  &    51    &    34.8  &    86.1  &    0.09  &     1.2  &     0.7  &     0.5  &     1.8  &  U-STA  \\
5   &     04:19:29.7   &  +37:22:50  &    52    &    23.5  &    17.6  &    0.07  &     1.1  &     0.5  &     0.5  &     1.4  &  C-PRE  \\
6   &     04:19:36.4   &  +37:29:05  &    27.8  &    18.2  &   128.5  &    0.03  &     0.8  &     0.3  &     0.2  &     0.6  &  C-PRE  \\
7   &     04:19:36.7   &  +37:28:46  &    26.3  &    18.2  &   133.8  &    0.03  &     0.7  &     0.3  &     0.1  &     0.6  &  U-STA  \\
8   &     04:19:37.2   &  +37:26:48  &    52.3  &    36.3  &   102    &    0.1   &     2.2  &     0.8  &     1.4  &     1.9  &  R-PRE  \\
9   &     04:19:39.8   &  +37:28:59  &    48.8  &    35.5  &    85.7  &    0.09  &     1.2  &     0.4  &     1.2  &     1.8  &  R-PRE  \\
10  &     04:19:49.5   &  +37:24:01  &    34.5  &    24.3  &   151.2  &    0.05  &     0.8  &     0.5  &     0.2  &     1.1  &  U-STA  \\
11  &     04:19:57.5   &  +37:31:06  &    43.4  &    23.4  &    95    &    0.06  &     1.1  &     0.3  &     0.3  &     1.3  &  C-PRE  \\
12  &     04:20:03.7   &  +37:31:24  &    50.6  &    18.4  &    52.4  &    0.06  &     1.5  &     0.5  &     4.4  &     1.2  &  R-PRE  \\
13  &     04:20:14.8   &  +37:31:53  &    49.2  &    32.1  &    84.9  &    0.09  &     2.9  &     1.1  &     1.1  &     1.7  &  R-PRE  \\
14  &     04:20:26.4   &  +37:10:25  &    37.7  &    25.1  &    49.6  &    0.06  &     1.2  &     0.8  &     0.2  &     1.2  &  U-STA  \\
15  &     04:20:30.6   &  +37:33:31  &    46.3  &    21.8  &    85.7  &    0.06  &     0.6  &     0.3  &     0.1  &     1.3  &  U-STA  \\
16  &     04:20:33.3   &  +37:12:09  &    44.1  &    25.2  &     0.1  &    0.07  &     0.4  &     0.3  &     0.1  &     1.3  &  U-STA  \\
17  &     04:20:34.9   &  +37:15:38  &    40.7  &    20.6  &     2.6  &    0.05  &     0.2  &     0.1  &    0.06  &     1.1  &  U-STA  \\
18  &     04:20:42.1   &  +37:33:32  &    31.4  &    21.9  &    83.9  &    0.05  &     1.4  &     0.8  &     0.6  &     0.9  &  R-PRE  \\
19  &     04:20:45.5   &  +37:29:49  &    38.1  &    25.6  &    57.9  &    0.06  &     1.1  &     0.6  &     0.3  &     1.2  &  U-STA  \\
20  &     04:20:47.4   &  +37:30:38  &    44.5  &    35.9  &    38.5  &    0.09  &     1.5  &     0.8  &     2.2  &     1.7  &  R-PRE  \\
21  &     04:20:47.8   &  +37:17:16  &    77.5  &    26.5  &    40.4  &    0.1   &     0.7  &     0.3  &     0.5  &     2    &  U-STA  \\
22  &     04:20:53.9   &  +37:33:12  &    54.3  &    24.8  &    94.4  &    0.08  &     0.8  &     0.3  &     0.3  &     1.5  &  U-STA  \\
23  &     04:21:00.6   &  +37:32:41  &    31.1  &    20.6  &    55.6  &    0.04  &     1.5  &     0.4  &     0.2  &     0.8  &  C-PRE  \\
24  &     04:21:00.9   &  +37:31:19  &    28.6  &    20.6  &    97.6  &    0.04  &     0.5  &     0.2  &    0.04  &     0.8  &  U-STA  \\
25  &     04:21:01.2   &  +37:20:50  &    53.6  &    31.9  &     4.8  &    0.09  &     4.8  &     1.6  &     5.8  &     1.8  &  R-PRE  \\
26  &     04:21:02.8   &  +37:34:24  &    43.6  &    31.7  &     5.4  &    0.08  &     0.9  &     0.4  &     0.2  &     1.6  &  U-STA  \\
27  &     04:21:05.8   &  +37:24:12  &    42.3  &    33.6  &    33.3  &    0.08  &     1.4  &     0.6  &     0.5  &     1.6  &  C-PRE  \\
28  &     04:21:07.3   &  +37:25:20  &    52.3  &    30.5  &    37.1  &    0.09  &     2.2  &     0.8  &     1.3  &     1.7  &  R-PRE  \\
29  &     04:21:14.5   &  +37:37:48  &    46.2  &    18.5  &   169.9  &    0.06  &     6.3  &     1.4  &     0.9  &     1.1  &  R-PRE  \\
30  &     04:21:14.6   &  +37:36:37  &    41.2  &    20    &   164.1  &    0.05  &     5    &     0.9  &     0.9  &     1.1  &  R-PRE  \\
31  &     04:21:15.5   &  +37:31:49  &    23.8  &    18.2  &     7.8  &    0.02  &     0.9  &     0.2  &     0.2  &     0.5  &  C-PRE  \\
32  &     04:21:17.4   &  +37:35:38  &    37.8  &    20    &   116.5  &    0.05  &     4.3  &     0.5  &     0.7  &     1    &  R-PRE  \\
33  &     04:21:18.2   &  +37:33:44  &    37.6  &    19.8  &     4.8  &    0.05  &    10.9  &     1.1  &     2.3  &     1    &  R-PRE  \\
34  &     04:21:20.7   &  +37:34:51  &    39.1  &    20.5  &   165.8  &    0.05  &     2.4  &     0.3  &     0.5  &     1    &  R-PRE  \\
35  &     04:21:23.3   &  +37:32:20  &    22.4  &    18.7  &    90.3  &    0.02  &     0.7  &     0.1  &    0.05  &     0.5  &  U-STA  \\
36  &     04:21:24.4   &  +37:36:31  &    53    &    21.9  &    52.4  &    0.07  &     0.9  &     0.3  &     0.4  &     1.4  &  C-PRE  \\
37  &     04:21:26.2   &  +37:36:38  &    22.5  &    18.2  &    41.9  &    0.02  &     0.5  &     0.2  &    0.07  &     0.4  &  U-STA  \\
38  &     04:21:29.9   &  +37:33:24  &    51.8  &    42.8  &    87.7  &    0.11  &     2.4  &     0.5  &     0.8  &     2.1  &  C-PRE  \\
39  &     04:21:32.7   &  +37:42:23  &    50.4  &    36.4  &     5.4  &    0.09  &     0.6  &     0.5  &     0.2  &     1.9  &  U-STA  \\
40  &     04:21:33.7   &  +37:32:54  &    34.1  &    18.2  &    92.7  &    0.04  &     3.6  &     0.4  &     0.6  &     0.8  &  R-PRE  \\
41  &     04:21:35.6   &  +37:36:22  &    26.6  &    18.2  &   138.4  &    0.03  &     0.8  &     0.1  &     0.4  &     0.6  &  R-PRE  \\
42  &     04:21:37.4   &  +37:37:43  &    33.3  &    23    &    69.2  &    0.05  &     1.1  &     0.3  &     0.1  &     1    &  U-STA  \\
43  &     04:21:37.7   &  +37:35:21  &    32.5  &    18.2  &   157.4  &    0.04  &     5.6  &     0.6  &     2.8  &     0.8  &  R-PRE  \\
44  &     04:21:38.5   &  +37:34:37  &    24.1  &    18.2  &     3    &    0.03  &    18.7  &     1.9  &     3.8  &     0.5  &  PRO  \\
45  &     04:21:39.2   &  +37:32:31  &    45.9  &    18.3  &   136.4  &    0.05  &     4.3  &     0.5  &     0.3  &     1.1  &  C-PRE  \\
46  &     04:21:41.1   &  +37:33:58  &    20.3  &    18.7  &   130.3  &    0.02  &    15.3  &     1.2  &     1.9  &     0.3  &  PRO  \\
47  &     04:21:41.3   &  +37:37:16  &    27.2  &    22.1  &   116.7  &    0.04  &     0.9  &     0.2  &    0.1   &     0.8  &  U-STA  \\
48  &     04:21:45.2   &  +37:32:15  &    31    &    23.2  &     2.3  &    0.05  &     1.7  &     0.2  &     0.6  &     1    &  R-PRE  \\
49  &     04:21:45.5   &  +37:33:19  &    36.6  &    18.2  &   128.2  &    0.04  &     9    &     0.9  &     2.9  &     0.9  &  R-PRE  \\
50  &     04:21:50.6   &  +37:32:32  &    28.7  &    18.2  &   163.8  &    0.03  &     1.3  &     0.2  &     0.5  &     0.7  &  R-PRE  \\
51  &     04:21:51.2   &  +37:32:03  &    22.6  &    18.2  &   164.9  &    0.02  &     1.4  &     0.3  &     0.3  &     0.4  &  R-PRE  \\
52  &     04:21:55     &  +37:31:43  &    27.6  &    22.7  &   133.5  &    0.04  &     1    &     0.3  &     0.2  &     0.8  &  C-PRE  \\
53  &     04:21:55.2   &  +37:38:04  &    43.8  &    18.2  &    22.6  &    0.05  &     0.5  &     0.3  &     0.1  &     1    &  U-STA  \\
54  &     04:21:57.7   &  +37:20:10  &    59.7  &    29    &    14.4  &    0.09  &     0.5  &     0.3  &     0.2  &     1.8  &  U-STA  \\
55  &     04:22:02.8   &  +37:34:17  &    34.2  &    18.2  &   111.3  &    0.04  &     0.7  &     0.5  &     0.2  &     0.8  &  U-STA  \\
56  &     04:22:07.4   &  +37:16:49  &    53.6  &    32    &   166.5  &    0.09  &     0.9  &     0.5  &     0.3  &     1.8  &  U-STA  \\
57  &     04:22:09.7   &  +37:14:39  &    45.8  &    35.7  &    90.2  &    0.09  &     1    &     0.5  &     0.3  &     1.7  &  U-STA  \\
   \hline                                
\end{tabular}
\tablefoot{These cores were identified and classified using the standard HGBS procedure described by \citet{Konyves2015}.
\textbf{RA} and \textbf{DEC} are the centroid equatorial coordinates of the cores. The cores are sorted from west to east. 
\bm{${H}_{\rm L}$} and \bm{${H}_{\rm S}$} are the major and minor axes 
of the elliptical Gaussian source that was fitted to each core 
 in the {\it Herschel} high-resolution ($18.2${\arcsec}) column density map.
\textbf{PA} is the position angle of the major axis (measured east of north). 
$\boldsymbol{R_{\rm dec}}$ is the deconvolved core radius. 
$\boldsymbol{N_{\rm H_{2}}^{\rm p}}$ is the peak column density. 
$\boldsymbol{N^{\rm p}_{\rm H_{2}}/N^{\rm bg}_{\rm H_{2}}}$ is ratio of peak to background column density. 
$\boldsymbol{M_{\rm core}}$ is the core mass estimated from SED fitting. 
When a core is protostellar, $\boldsymbol{M_{\rm core}}$ is the protostellar envelope mass. 
$\boldsymbol{M_{\rm BE}}$ is the critical Bonnor-Ebert (BE) mass. 
\textbf{R-PRE} stands for robust prestellar core,  \textbf{C-PRE} for candidate prestellar core, 
\textbf{U-STA} for  unbound starless core, and \textbf{PRO} for protostellar core. 
There are 24 U-STA cores, 11 C-PRE cores, 20 R-PRE cores, and 2 PRO cores in total.}   
\end{table*}

\end{appendix}
\end{document}